\documentclass[trackchanges]{aastex63}
\usepackage{makecell}
\usepackage{float}
\usepackage{footmisc}
\usepackage{bibentry}
\usepackage{wasysym}
\shorttitle{Eldest X-Ray Supernovae}
\shortauthors{Ramakrishnan et al.}

\begin{document}

\title{From Supernova to Remnant: Tracking the Evolution of the Oldest Known X-ray Supernovae}

\correspondingauthor{Vikram V. Dwarkadas}
\email{vikram@astro.uchicago.edu}




\author[0000-0002-9176-7252]{Vandana Ramakrishnan}
\affiliation{Department of Astronomy and Astrophysics, University of
  Chicago\\ 5640 S Ellis Ave, Chicago, IL 60637}
\affiliation{Current Address: Department of Physics and Astronomy, Purdue University \\
525 Northwestern Ave, West Lafayette, IN 47907}

\author[0000-0002-4661-7001]{Vikram V. Dwarkadas}
\affiliation{Department of Astronomy and Astrophysics, University of
  Chicago\\ 5640 S Ellis Ave, Chicago, IL 60637}




\begin{abstract}
Core-collapse supernovae (SNe) expand into a medium created by winds from the pre-SN progenitor. The SN explosion and resulting shock wave(s) heat up the surrounding plasma, giving rise to thermal X-ray emission, which depends on the density of the emitting material. Tracking the variation of the X-ray luminosity over long periods of time thus allows for investigation of the kinematics of the SN shock waves, the structure of the surrounding medium, and the nature of the progenitor star.  In this paper X-ray observations of five of the oldest known X-ray supernovae -- SN 1970G, SN 1968D, SN 1959D, SN 1957D and SN 1941C -- are analyzed, with the aim of reconstructing their light curves over several decades. For those supernovae for which we can extract multi-epoch data, the X-ray luminosity {appears to} decline with time, although with large error bars. No increase in the X-ray emission from SN 1970G is found at later epochs, contrary to previous reports. All five SNe show X-ray luminosities that are of comparable magnitude. We compare the late-time X-ray luminosities of these SNe to those of  supernova remnants (SNRs) in the Galaxy which are a few hundred years old, and find that when the {tentative decline is taken into account, the luminosity of the old SNe studied herein could fall below the luminosity of some of the younger SNRs within a few hundred years.} However, the X-ray luminosity should begin to increase as the SNe expand in the Sedov phase, thus reaching that of the observed SNRs.

\end{abstract}

\section{Introduction} 
\label{sec:intro}

The core-collapse of a massive star gives rise to an extremely
fast-moving forward shock that sweeps up the ambient medium, and a
reverse shock that propagates back into the ejecta (note that even the
reverse shock propagates outwards initially in the frame of a viewer
on earth, although it is moving inwards in the frame fixed to the
outer shock \citep[see for example][]{dc98}). In core-collapse supernovae (SNe), the
ambient medium is generally formed by mass-loss from the progenitor
star, and is referred to as the circumstellar medium (CSM). As the
shocks propagate outwards, they heat the material to temperatures
$\sim$ 10$^6$ - 10$^9$ K. The hot gas will primarily emit in
X-rays. Thermal X-ray emission depends (among other factors) on the
density of the shocked medium. As a result, observations of the
thermal X-ray emission from a supernova can be a rich source of
information on the density structure of the supernova ejecta and the
CSM. Assuming power-law profiles for the ejecta and the CSM,
\citet{Chevalier1982} derived a self-similar solution for the
interaction of the supernova shock wave with the CSM, which describes
the evolution of the forward and reverse shock waves. Starting from
the self-similar solution for the SN evolution, the X-ray emission
from supernovae can also be described in a semi-analytic
fashion. \citet{Chevalier2017} provide an exhaustive review on the
X-ray and radio evolution of young core-collapse SNe.

The CSM, created by the wind from the progenitor star, encodes
information regarding the mass loss rate of the progenitor at various
points in time. As the shock wave expands outwards, the X-ray emission
traces the wind evolution, and thus contains information regarding the
wind properties before the star exploded.  Since the velocity of the
shock wave far exceeds that of the progenitor wind in most cases, the
shock wave can trace several decades of wind evolution in a very short
time. For example, in the case of a red supergiant progenitor, the
wind velocity is of order $\sim$ 10 km s$^{-1}$, whereas the shock
velocity is initially of order $\sim$ 10,000 km s$^{-1}$. In
propagating for one year, the shock wave traces $\sim$ 1,000 years of
wind evolution. The shock velocity will decrease with time, but the
X-ray emission 10 years after the explosion can still provide
information on the progenitor mass loss several thousand years before
the explosion. The greater the duration over which the X-ray light
curve is constructed, the greater the lookback time. Thus extremely
long X-ray lightcurves, implying X-ray observations of older SNe, are
highly desirable.

The X-ray light curves of old supernovae make it possible to bridge
the gap between supernovae and supernova remnants  {(SNRs)}. A large number of
Galactic SNRs have been observed in X-rays (Cas A,
Kepler, and Tycho are some of the most well known). Unfortunately
there has been no optically observed SN in our Galaxy since Cas A,
almost 340 years ago, with the nearest well-observed SN with
modern-day telescopes being SN 1987A in the Large Magellanic Cloud (LMC). Further, there exist
only around a dozen known X-ray emitting supernovae from before 1990,
i.e. older than $\sim$ 30 years, whereas the youngest remnant in our
Galaxy, G1.9+0.3, has an age of $\sim$ 100 years
\citep{Reynolds2008}. Therefore, although models (such as
\citet{Chevalier1982}) exist that attempt to describe the evolution of
a young SN to a remnant, the actual transition to the remnant stage,
the expansion and deceleration of the shock waves, and the evolution
of the X-ray emission over the first few hundred years, is largely
uncharted territory. This is due to both the fact that many SNe decay
and are unobservable after a couple of decades, and that young SNe are
not continually followed in X-rays over several decades, due to
pressure on available resources.

Following the first detection of X-ray emission from a supernova
(SN 1980K, \citet{SN1980k}), X-ray supernovae have been found at an
ever-increasing rate thanks to the rapid progress of astronomical
facilities. In 1995, 5 X-ray supernovae were known
\citep{Schlegel1995}; in 2003, $\sim$ 30 \citep{Immler2003}. At
present, $>$ 60 X-ray supernovae are known
\citep{Dwarkadas_2012,Dwarkadas2014,rd17}. In particular, the sub-arcsecond
resolution and excellent point source sensitivity of the ACIS
instrument\footnote{http://cxc.harvard.edu/cdo/about\_chandra} of the
\textit{Chandra} X-ray Observatory, makes it possible to detect faint
emission from old supernovae (or young remnants, depending on the
definition) above the background. SN 1941C was detected with Chandra in
2001 \citep{Soria_2008}, at the age of 61 years! Archival data can be
used to construct the light curves of supernovae out to late times,
providing an opportunity to study the transition from SN to SNR.

In this paper we present and discuss archival \textit{Chandra} X-ray
observations of five of the oldest known core-collapse X-ray SNe - SN 1941C, SN 1957D,
SN 1959D, SN 1968D and SN 1970G.\footnote{All of these SNe are optically observed, although their actual classification may be unknown. Nevertheless, all SNe detected in X-rays have been core-collapse SNe \citep{Dwarkadas_2012,rd17}, except for an unusual one that is classified as a Type Ia-CSM \citep{bocheneketal18}. Searches for X-ray emission from all other Type Ia SNe have turned out to be negative. Therefore we feel it safe to assert that all the SNe studied herein have a core-collapse origin.} The extraction and analysis of spectra
obtained from archival observations is discussed in Section
\ref{sec:analysis}. The X-ray fluxes and luminosities for all the
supernovae, and the light curves for SN 1957D, SN 1968D and SN 1970G, are
presented in Section \ref{sec:results}. These light curves extend
further out than those from any previous study. We discuss each SN,
and compare our results with past observations when applicable, in
Section \ref{subsec:individual}. The implications of our results are
discussed in Section \ref{subsec:light_curves}.Summary
of our work, and conclusions, are  presented in Section \ref{sec:conclusion}.

\section{Data Analysis} \label{sec:analysis}
We have analyzed all archival and publicly available \textit{Chandra}
observations of SN 1941C, SN 1957D, SN 1959D, SN 1968D and SN
1970G. The data were reduced and reprocessed using \textsc{ciao}
software version 4.10 \citep{ciao} and \textsc{caldb} database version
4.8.0. Spectral fitting on the reduced data was done using
\textsc{sherpa} software. In the absence of further information, and due to the low count rates in all of the observations, it was necessary to make some assumptions in order to carry out fitting. Firstly, the column density was fixed at the Galactic value in all cases {except for SN 1959D}. For most SNe {whose host galaxies are viewed face-on}, the major contributions to the absorbing column come from the material along the line of sight in our Galaxy, and the material around the SN itself. {Of the five SNe studied herein, 4 of them occur in galaxies which are nearly face-on or have low inclination angles. The exception is SN 1959D in the galaxy NGC 7331, which has a high inclination angle and is viewed nearly edge-on \citep{Kent85}, and for which an extra contribution to the column density from the host galaxy must be taken into account}. Core-collapse SNe evolve in the winds of their progenitor stars, whose density would be expected to drop as r$^{-2}$ if the wind parameters were constant. Consequently, as the SN shock expands, it crosses the higher-density material surrounding the SN, and subsequently expands into lower-density material that does not contribute much to the column density. This is seen, for example, in SN 1993J, where the column density became consistent with the Galactic value after around 2000 days \citep{sn1993j}. {Considering} the age of the SNe studied here, and the low statistics, the assumption of the Galactic column density as being the only contribution to the absorption seems a reasonable one {for all the SNe except SN 1959D.} Second, most core-collapse SNe, except for those of Type Ib/c and Type IIP, are found to show thermal emission, with few exceptions. The IIPs have the lowest luminosity amongst all the SN classes \citep{Dwarkadas_2012, Dwarkadas2014}, and are unlikely to be seen at late times. Since we are unable to determine the precise spectral model due to low count rates, fitting the spectra with a thermal (MEKAL) model in all cases (except SN 1957D) seemed prudent. We note that most previous authors who have dealt with these SNe have similarly used thermal models \citep{Immler_2005, Soria_2008}. In the case of SN 1970G, a non-thermal (powerlaw) model was tried, but the thermal model is found to be favored (see Section \ref{subsec:individual}). Finally, the ionization temperature of the shocked plasma may be lower than the thermal temperature for many young SNRs (IC443, W49B) i.e. the SNe are not in ionization equilibrium, and non-equilibrium ionization (NEI) models may perhaps be preferable, depending on the density. However NEI models are difficult to constrain with low statistics data. Therefore, past descriptions of the SNe in this paper \citep{Immler_2005, Soria_2008} have employed models in ionization equilibrium, such as the MEKAL model, and we do the same in this paper, as the paucity of counts does not allow for more detailed fits in most cases. It is worth mentioning that most of the NEI models available in XSPEC, although sometimes used for young SNe, contain approximations not readily applicable to young SNe, such as planar shocks (vpshock), a Sedov profile (vsedov) or constant density (NEI).  

In this work the absorption component was represented by the \textit{xstbabs} model, although we have found that using the \textit{xsphabs} model did not significantly affect any of the derived fluxes. The unabsorbed fluxes are listed in the 0.3-2 keV band as well as the 0.3-8 keV band, to enable comparisons of the calculated fluxes with past works. Data reduction and analysis for
each individual SN is discussed in detail below.

\subsection{SN 1970G} \label{subsec:sn1970g}
SN 1970G in the galaxy M101 was first observed on 1970 July 30
\citep{1970g_discovery}, by researchers at the Konkoly Observatory in
Budapest \citep{detre74}. It was the first radio supernova ever to be
observed \citep{gottesmanetal72}, and was recovered again at radio
wavelengths in 1990.

\paragraph{ROSAT} SN 1970G was first detected in X-rays in a 1 Ms {\it Chandra} observation of the host galaxy M101, following which an archival data search revealed that a source at the same position also existed in
{\it ROSAT} data \citep{Immler_2005}. The presence of a nearby HII region NGC 5455 $\approx5''$ away complicated the derivation of the flux from ROSAT and {\it XMM-Newton}, with the latter yielding only upper limits. However, using the exact spatial position from {\it Chandra},
\citet{Immler_2005} managed to recover count rates and fluxes from data taken with the \textit{ROSAT} PSPC instrument and \textit{ROSAT} HRI instrument. The count rates for the \textit{ROSAT} observations given in
\citet{Immler_2005} were used in this paper to obtain fluxes in the
relevant bands, using the \textit{Chandra} \textsc{pimms} tool \footnote{http://cxc.harvard.edu/toolkit/pimms.jsp}. Anticipating the results of our spectral fitting of \textit{Chandra} data from 2004 (below), a plasma MEKAL model with a temperature of 0.61 keV, and Galactic column density towards the remnant of 1.17 $\times$ 10$^{20}$
cm$^{-2}$ \citep{DL_nH}, was used to derive the unabsorbed flux. The
results did not change significantly with the assumption of a higher
temperature of 1.08 keV. The unabsorbed fluxes were calculated in the
0.3-2 keV band to enable comparisons with \citet{Immler_2005}, as well as in
the 0.3-8 keV band to enable uniform comparisons with other SNe.
\newline

\paragraph{Chandra} Three publicly available \textit{Chandra} datasets,
taken using ACIS-S in 2004, 2011 and 2017 respectively, were used for the analysis. The details of these archival observations are listed in Table \ref{tab:sn1970g_obs}. 

\begin{figure}
    \centering
    \includegraphics[width=0.7\linewidth]{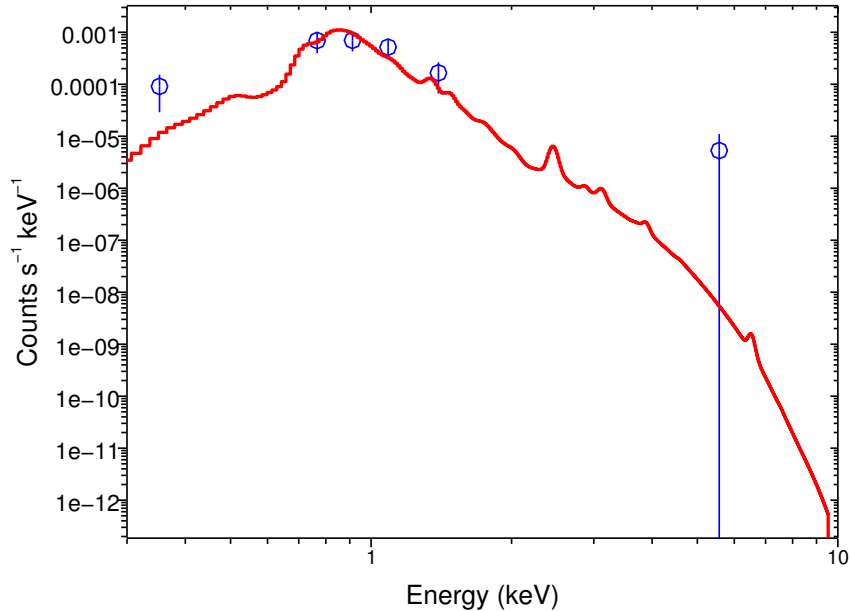}
    \caption{The spectrum and best fit model extracted from the 2004 observation of SN 1970G. The spectrum was grouped by 15 counts and fitted after background subtraction.}
    \label{fig:sn1970g_spec_1}
\end{figure}

\begin{figure}
    \centering
    \includegraphics[width=0.7\linewidth]{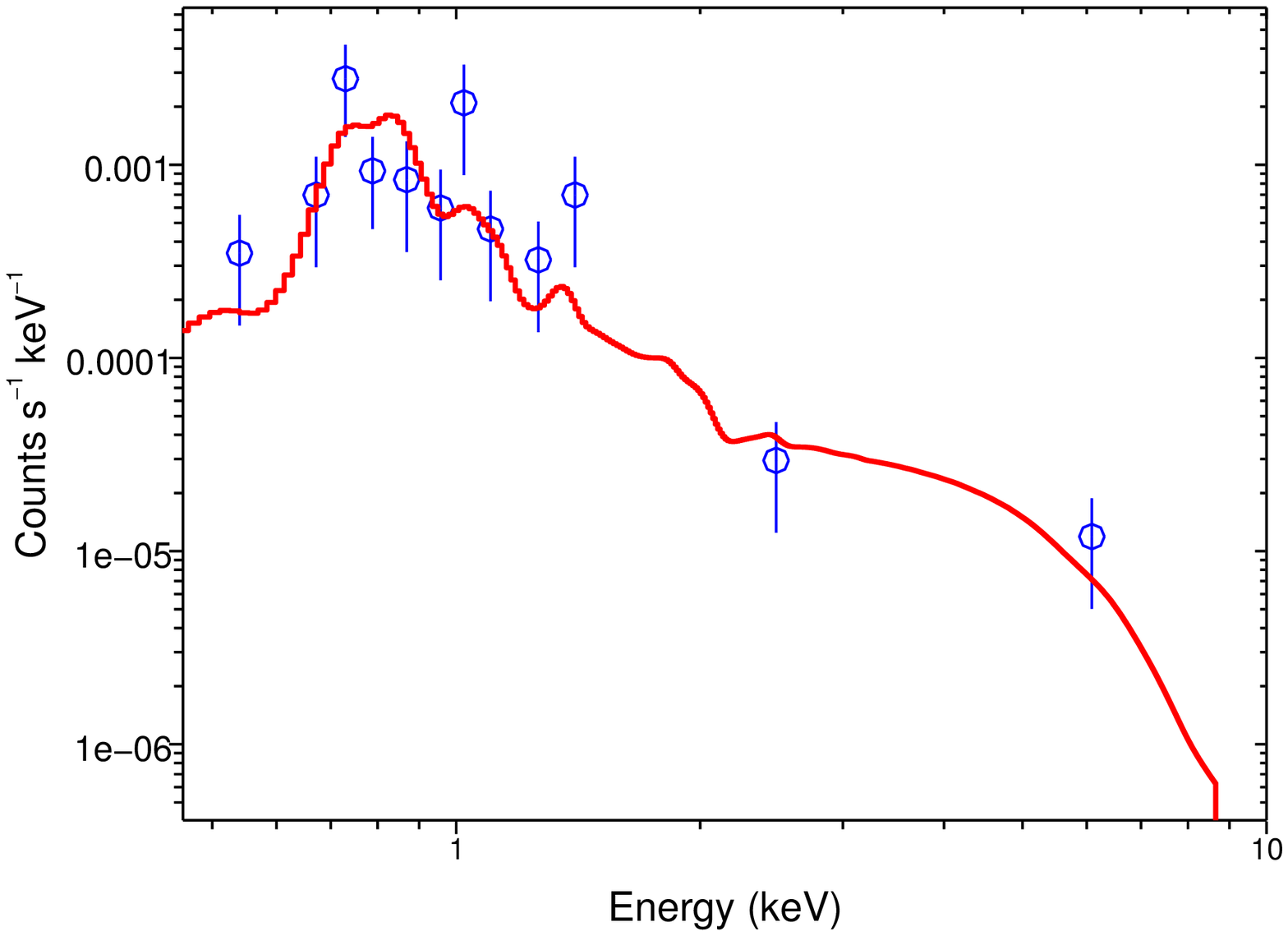}
    \caption{The spectrum and best fit model extracted from the 2011 observation of SN 1970G. Note that while the spectrum shown here was grouped by 3 counts for ease of visibility, the model was produced by fitting the ungrouped spectrum}
    \label{fig:sn1970g_spec_2}
\end{figure}

The data from 2004 and 2011 were reduced using standard
procedures. The source and background spectra were extracted using a
4-arcsecond circular source region centered on the source coordinates
(which includes 90\% of the encircled energy), and a 4-arcsecond
circular background region near the source. The source coordinates
were taken to be those given in the \textsc{SIMBAD} astronomical
database \citep{SIMBAD}. The spectra extracted from the 2004
observations were co-added using the \textsc{ciao} script
\textit{combine\_spectra}. Given the large number of counts, we
grouped the co-added spectrum into bins of 15 counts, subtracted the
background, and fit the spectrum, using a \textit{chi2gehrels}
statistic to analyze the goodness of the fit. The 2011 spectrum had
a much lower number of counts, and the \textit{chi2gehrels} statistic
would not have been appropriate. The source and background
spectra were simultaneously fit using the Cash (\textit{cstat})
statistic. Both the co-added spectrum from the 2004 observations and
the spectrum from the 2011 observation were fit with an absorbed
thermal plasma \textit{vmekal} model, with the absorption component
being described by the \textit{xstbabs} model. The column density was
frozen at the Galactic value (1.17 $\times$ 10$^{20}$
cm$^{-2}$). Spectral fitting of the 2004 and 2011 spectra resulted in
temperatures of 0.63 keV and 0.44 keV respectively (see Table
\ref{tab:sn1970g}). The spectrum and best fit model for the 2004 and 2011 observations of SN 1970G are shown in Figures \ref{fig:sn1970g_spec_1} and \ref{fig:sn1970g_spec_2}.

The 2017 spectrum had only 12 source counts in the 0.5 - 8 keV band
(with a 4-arcsecond source region). In order to derive a flux, we used
the \textit{srcflux} command from \textsc{ciao}, which provided us
with the count rate at the source position. The count rate was
converted to a flux value using the \textit{Chandra} \textsc{pimms}
tool. In accordance with the fit for the 2011 spectrum, the flux was
found using a plasma MEKAL model with temperature 0.44 keV and
Galactic column density.

The number of days since discovery, count rates, temperature and flux
at all epochs for SN 1970G are presented in Table \ref{tab:sn1970g}.

\renewcommand{\thefootnote}{\arabic{footnote}}

\begin{table}
    \centering
    \begin{tabular}{c c c c}
    \hline
         \colhead{Year} & \colhead{Observation ID} & \colhead{Date} & \colhead{Total Exposure (ks)} \\
    \hline     
         2004 & 5337 & 2004 Jul 5 & 10 \\ 
              & 5338 & 2004 Jul 6 & 29 \\
              & 5339 & 2004 Jul 7 & 14 \\
              & 5340 & 2004 Jul 8 & 54 \\
              & 4734 & 2004 Jul 11 & 35 \\
              & & & Total 142 \\
         2011 & 14341 & 2011 Aug 27 & 49 \\
         2017 & 19304 & 2017 Nov 8 & 37 \\
    \hline     
    \end{tabular}
    \caption{Details of archival \textit{Chandra} observations of SN1970G. }
    \label{tab:sn1970g_obs}
\end{table}

\begin{table}
    \centering
    \begin{tabular}{c c c c c c c c}
        \hline 
         \colhead{Instrument} & \colhead{Date of} & \colhead{Days since} & \colhead{Count rate} &  \colhead{kT (keV)} & \colhead{Reduced} & \colhead{Flux (0.3 - 2 keV)} & \colhead{Flux (0.3 - 8 keV)} \\
         \colhead{} & \colhead{observation} & \colhead{discovery} & \colhead{($\rm 10^{-4}~counts~s^{-1}$)} & \colhead{} & \colhead{statistic} & \colhead{($\rm 10^{-15}~erg~cm^{-2}~s^{-1}$)} & \colhead{($\rm 10^{-15}~erg~cm^{-2}~s^{-1}$)} \\
        \hline
         ROSAT & 1991 Jun 8-9 & 7618 & 9.0 $\pm$ 2.1 & 0.61 \footnote{\label{note1} Assumed value} & - &  7.09 $\pm$ 1.65 & 7.23 $\pm$ 1.69 \\
         (PSPC) & & & & & \\
         ROSAT & 1996 May 14 & 9516 & 1.8 $\pm$ 0.4 & 0.61$\rm^{\ref{note1}}$ & - & 4.47 $\pm$ 0.99 & 4.59 $\pm$ 1.02 \\
         (HRI) & - Nov 23 & & & & \\
         Chandra & 2004 Jul 5-11 & 12397 & 4.06 $\pm$ 0.86 & 0.63 $\pm$ 0.13 & 0.79 & 2.33 $\pm$ 0.59 & 2.35 $\pm$ 0.60 \\
         (ACIS-S) & & & & & & \\
         Chandra & 2011 Aug 27 & 15003 & 5.19 $\pm$ 1.46 & 0.44 $\pm$ 0.40 & 0.21 & 2.31 $\pm$ 1.23 & 2.41 $\pm$ 1.33 \\
         (ACIS-S) & & & & & \\
         Chandra & 2017 Nov 8 & 17268 & 0.40$^{+0.49}_{-0.31}$ & 0.43$\rm ^{\ref{note1}}$ & - & 0.41$^{+0.50}_{-0.32}$ & 0.42$^{+0.51}_{-0.32}$ \\ 
         (ACIS-S) & & & & & & \\
        \hline
    \end{tabular}
    \caption{X-ray data for SN 1970G, giving 1-$\sigma$ (68\%)
      error-bars where possible. The column density is fixed to the
      Galactic value of 1.17 $\times$ 10$^{20}$ cm$^{-2}$ }
    \label{tab:sn1970g}
\end{table}

\subsection{SN 1968D} \label{subsec:sn1968d}

SN 1968D is one of several optically discovered SNe in the galaxy NGC
6946, which is also the host galaxy of the first SN ever to be
detected in X-rays, SN 1980K. The supernova was discovered on 1968
February 29 \citep{1968d_discovery1,1968d_discovery2}. ROSAT observed
this region for over 36 ks in 1992, but the detection of SN 1968D
could not be confirmed \citep{schlegel94}.

\textit{Chandra} ACIS-S observations of NGC 6946 were carried out
several times in the last two decades, and we use these observations to construct the light curve of SN 1968D. Five sets of \textit{Chandra} observations were used, taken in 2001-02, 2004, 2012, 2016 and 2017 respectively. Details are given in Table \ref{tab:sn1968d_obs}.

\begin{figure}
    \centering
    \includegraphics[width=0.7\linewidth]{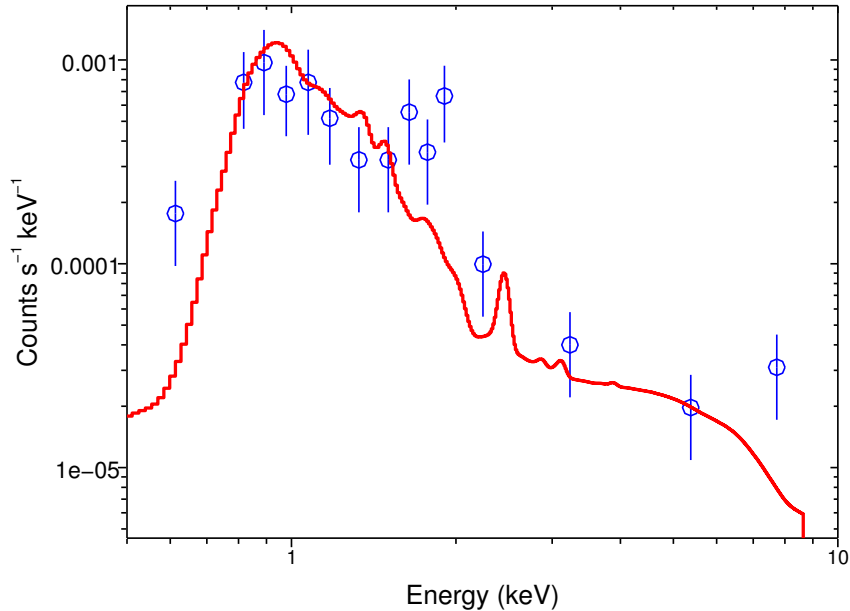}
    \caption{The spectrum and best fit model extracted from the 2001-02 observation of SN 1968D. Note that while the spectrum shown here was grouped by 5 counts for ease of visibility, the model was produced by fitting the ungrouped spectrum}
    \label{fig:sn1968d_spec_1}
\end{figure}

\begin{figure}
    \centering
    \includegraphics[width=0.7\linewidth]{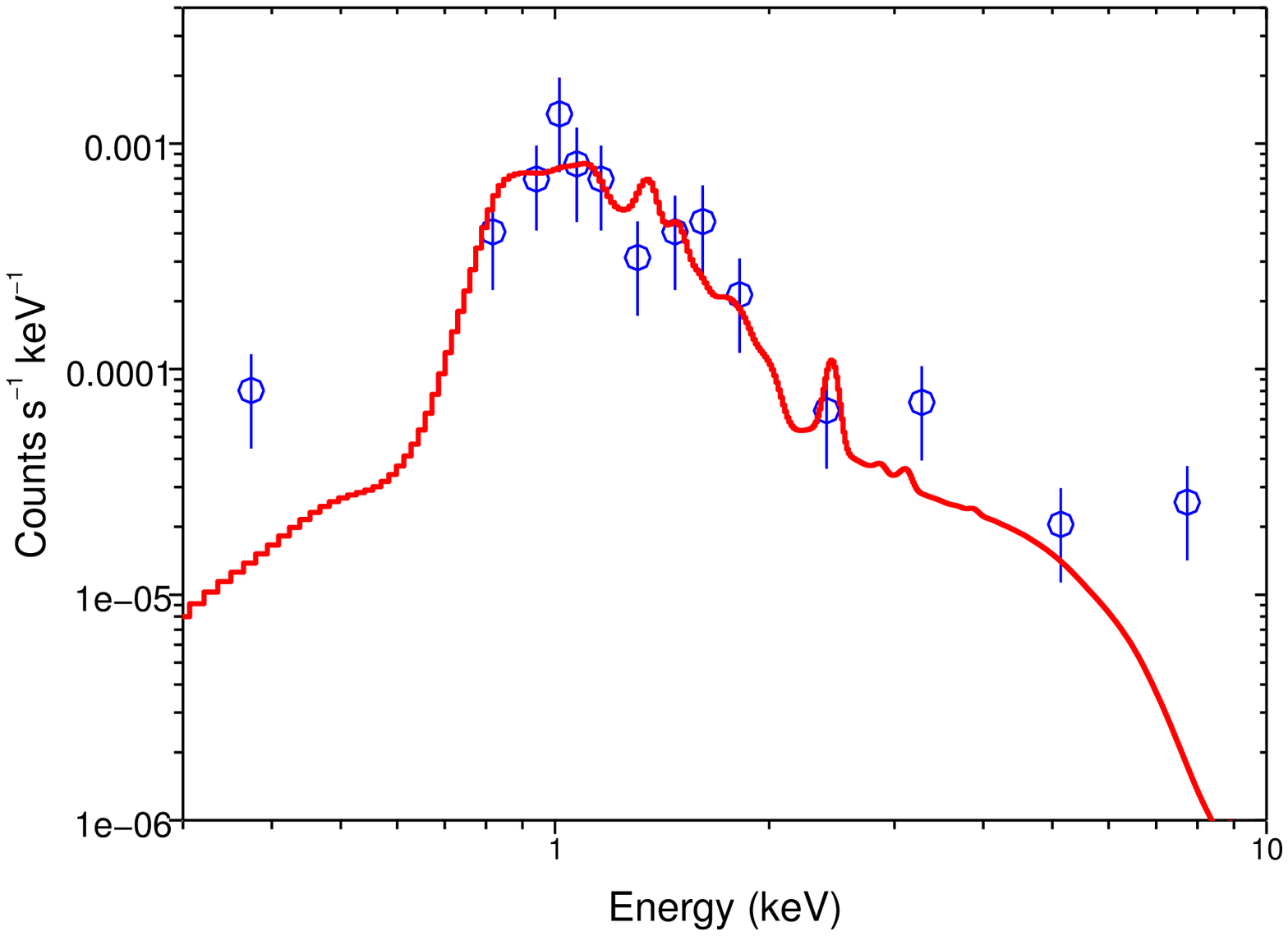}
    \caption{The spectrum and best fit model extracted from the 2004 observation of SN 1968D. Note that while the spectrum shown here was grouped by 5 counts for ease of visibility, the model was produced by fitting the ungrouped spectrum}
    \label{fig:sn1968d_spec_2}
\end{figure}

\begin{figure}
    \centering
    \includegraphics[width=0.7\linewidth]{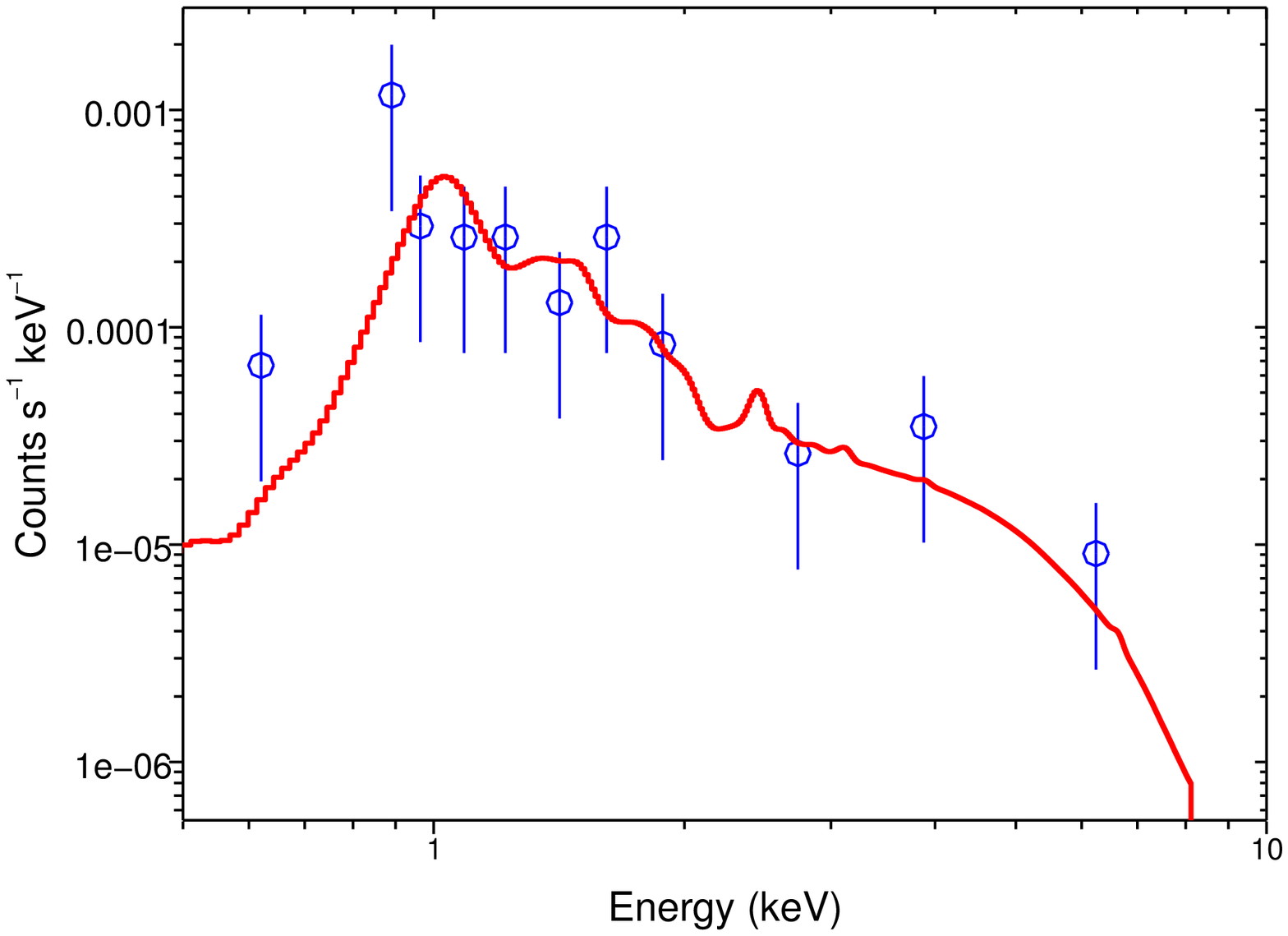}
    \caption{The spectrum and best fit model extracted from the 2016 observation of SN 1968D. Note that while the spectrum shown here was grouped by 2 counts for ease of visibility, the model was produced by fitting the ungrouped spectrum}
    \label{fig:sn1968d_spec_3}
\end{figure}

For each individual observation in the 2001-02, 2004 and 2016
datasets, the source and background spectra were extracted from a
4-arcsecond circular source region centered on the source coordinates
(from the SIMBAD astronomical database), and an annular background
region of inner radius 4 arcseconds and outer radius 8 arcseconds also
centered on the source. The individual spectra for each epoch were
co-added using the \textsc{ciao} script \textit{combine\_spectra}. The
co-added, ungrouped source and background spectral files were
simultaneously fit, using the \textit{cstat} statistic. Source spectra
were modeled with an absorbed thermal plasma \textit{vmekal} model,
using the \textit{xstbabs} model for the absorption component. The
column density was frozen at the Galactic value of 2.11 $\times$
10$^{21}$ cm$^{-2}$ \citep{DL_nH}. The fits resulted in a temperature
of 1.31 keV for the 2001-02 observations, 1.74 keV for the 2004
observations and 1.20 keV for the 2016 observations. The spectra and best fit models for these observations are shown in Figures \ref{fig:sn1968d_spec_1}, \ref{fig:sn1968d_spec_2} and \ref{fig:sn1968d_spec_3}. 

The 2012 and 2017 observations had a very low number of counts, 9 and
5 respectively, in the 0.3 - 8 keV band, assuming a 4-arcsecond source
region. This was clearly insufficient to allow for dividing the
spectra into bins and fitting. Instead, for each of these
observations, the \textit{srcflux} command was used to find the count
rate at the source location. The count rates were converted to flux
values using the Chandra \textsc{pimms} tool. On the basis of the fit
results from the 2016 epoch, a plasma MEKAL model with a temperature
of 1.22 keV and Galactic column density was used to compute the flux.

The flux estimates and other fitting parameters for SN 1968D are
presented in Table \ref{tab:sn1968d}.

\begin{table}
    \centering
    \begin{tabular}{c c c c}
    \hline
         \colhead{Year} & \colhead{Observation ID} & \colhead{Date} & \colhead{ Total Exposure (ks)} \\
    \hline
         2001 & 1043 & 2001 Sep 7 & 48 \\
         2002 & 4404 & 2002 Nov 25 & 30 \\
         2004 & 4631 & 2004 Oct 22 & 30 \\
              & 4632 & 2004 Nov 6 & 28 \\
              & 4633 & 2004 Dec 3 & 27 \\
              & & & Total 85 \\
         2012 & 13435 & 2012 May 21 & 20 \\
         2016 & 17878 & 2016 Sep 28 & 40 \\
              & 19887 & 2016 Sep 28 & 19 \\
              & & & Total 59 \\
         2017 & 19040 & 2017 Jun 11 & 10 \\
    \hline
    \end{tabular}
    \caption{Details of archival \textit{Chandra} observations of SN1968D }
    \label{tab:sn1968d_obs}
\end{table}

\begin{table}
    \centering
    \begin{tabular}{c c c c c c c c}
        \hline 
         \colhead{Instrument} & \colhead{Date of} &  \colhead{Days since} & \colhead{Count rate} &  \colhead{kT (keV)} & \colhead{Reduced} & \colhead{Flux (0.3 - 2 keV)} & \colhead{Flux (0.3 - 8 keV)} \\
         \colhead{} & \colhead{observation} & \colhead{discovery} & \colhead{($\rm 10^{-4}~counts~s^{-1}$)} & \colhead{} & \colhead{statistic} & \colhead{($\rm 10^{-15}~erg~cm^{-2}~s^{-1}$)} & \colhead{($\rm 10^{-15}~erg~cm^{-2}~s^{-1}$)} \\
        \hline
         Chandra & 2001 Sep 7 - & 12466 & 6.87 $\pm$ 1.37 & 1.31 $\pm$ 0.14 & 0.42 & 3.41 $\pm$ 0.95 & 4.07 $\pm$ 0.90 \\
         (ACIS-S) & 2002 Nov 25 & & & & & & \\
         Chandra & 2004 Oct 22 & 13406 & 4.47 $\pm$ 1.43 & 1.74$^{+1.09}_{-0.24}$ & 0.36 & 3.43 $\pm$ 0.49 & 4.69 $\pm$ 0.57 \\
         (ACIS-S) & - Dec 03 & & & & & & \\
         Chandra & 2012 May 21 & 16153 & 3.79 $\pm$ 1.58 & 1.22 \footnote{\label{note5}Assumed value} & - & 2.53 $\pm$ 1.05 & 3.02 $\pm$ 1.26 \\
         (ACIS-S) & & & & & & & \\
         Chandra & 2016 Sep 28 & 17744 & 2.33 $\pm$ 1.50 & 1.20$^{+0.26}_{-0.51}$ & 0.22 & 2.24 $\pm$ 1.17 & 2.68 $\pm$ 1.24 \\
         (ACIS-S) & & & & & & & \\
         Chandra & 2017 Jun 11 & 18000 & 0.74$^{+1.44}_{-0.73}$ & 1.22$^{\rm \ref{note5}}$ & - & 0.72$^{+1.40}_{-0.71}$ & 0.85$^{+1.65}_{-0.84}$  \\
         (ACIS-S) & & & & & & & \\
        \hline
    \end{tabular}
    \caption{X-ray data for SN 1968D, giving 1-$\sigma$ (68\%)
      error-bars where possible. The column density is fixed to the
      Galactic value of 2.11 $\times$ 10$^{21}$ cm$^{-2}$ }
    \label{tab:sn1968d}
\end{table}

\subsection{SN 1957D} 
\label{subsec:sn1957d}
M83 is another galaxy that, similar to NGC 6946, hosts many
intermediate age SNe, including one of the oldest {optically observed SNe}, SN 1923A. It also
hosts several X-ray SNe, including the bright SN 1983N, and SN
1957D. The latter is the only one of the 5 historical SNe in M83 that
is observed at optical \citep{longetal89}, radio \citep{cowanetal83}
and X-ray \citep{Long_2012} wavelengths. Little is known about the
early stages of this supernova; it was first observed in 1957
December, at a photographic magnitude $\leq$ 15
\citep{1957d_discovery}, but it was likely well past optical maximum
at this point. We have chosen to measure the time elapsed from 1957
June 15.

The light curve of SN 1957D presented herein was constructed using
publicly available {\it Chandra} data taken over 5 epochs. All
observations used the ACIS-S instrument, except the most recent one in
2014 which used ACIS-I. The observations were taken respectively in 2000, 2001, 2010, 2011 and 2014. The details of these archival observations are given in Table \ref{tab:sn1957d_obs}. 

Though the 2000 and 2001 data indicated a large number of counts in
the spectrum of SN 1957D, the same was true for the background
region. In the 2000 observation, in the 0.3 - 8 keV range, there were
29 source counts and 29 background counts. Furthermore, the source and
background spectra appeared to be very similar. The 2001 observation
had 8 source counts and 7 background counts in the 0.3-8 keV range,
and the source and background spectra were again similar. It appeared
highly unlikely that the supernova had been detected. For these two
observations, only an upper limit on the supernova flux could be
obtained. The count rates obtained using the \textsc{ciao} command
\textit{srcflux} were not used; instead the 3-sigma (99\%) upper
limits on the count rates obtained from \textit{srcflux} were
used. The count rates were converted to flux values using the
\textit{Chandra} \textsc{pimms} tool, assuming first a plasma MEKAL
model with Galactic column density (3.94 $\times$ 10$^{20}$ cm$^{-2}$,
\citet{DL_nH}) and a temperature of 3.43 keV; and then a powerlaw
model with Galactic column density and photon index 1.06. These values
of temperature and photon index were chosen based on the spectral fits
to the 2010 and 2011 datasets.
 
\begin{figure}
    \centering
    \includegraphics[width=0.7\linewidth]{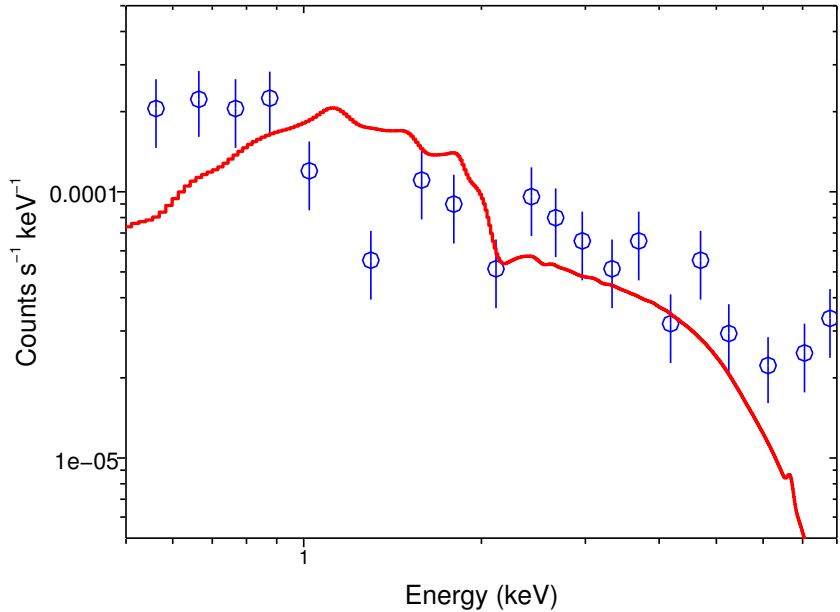}
    \caption{The spectrum and best fit thermal model extracted from the 2011 observations of SN 1957D. Note that while the spectrum shown here was grouped by 12 counts for ease of visibility, the model was produced by fitting the ungrouped spectrum.}
    \label{fig:sn1957d_spec_1}
\end{figure}

\begin{figure}
    \centering
    \includegraphics[width=0.7\linewidth]{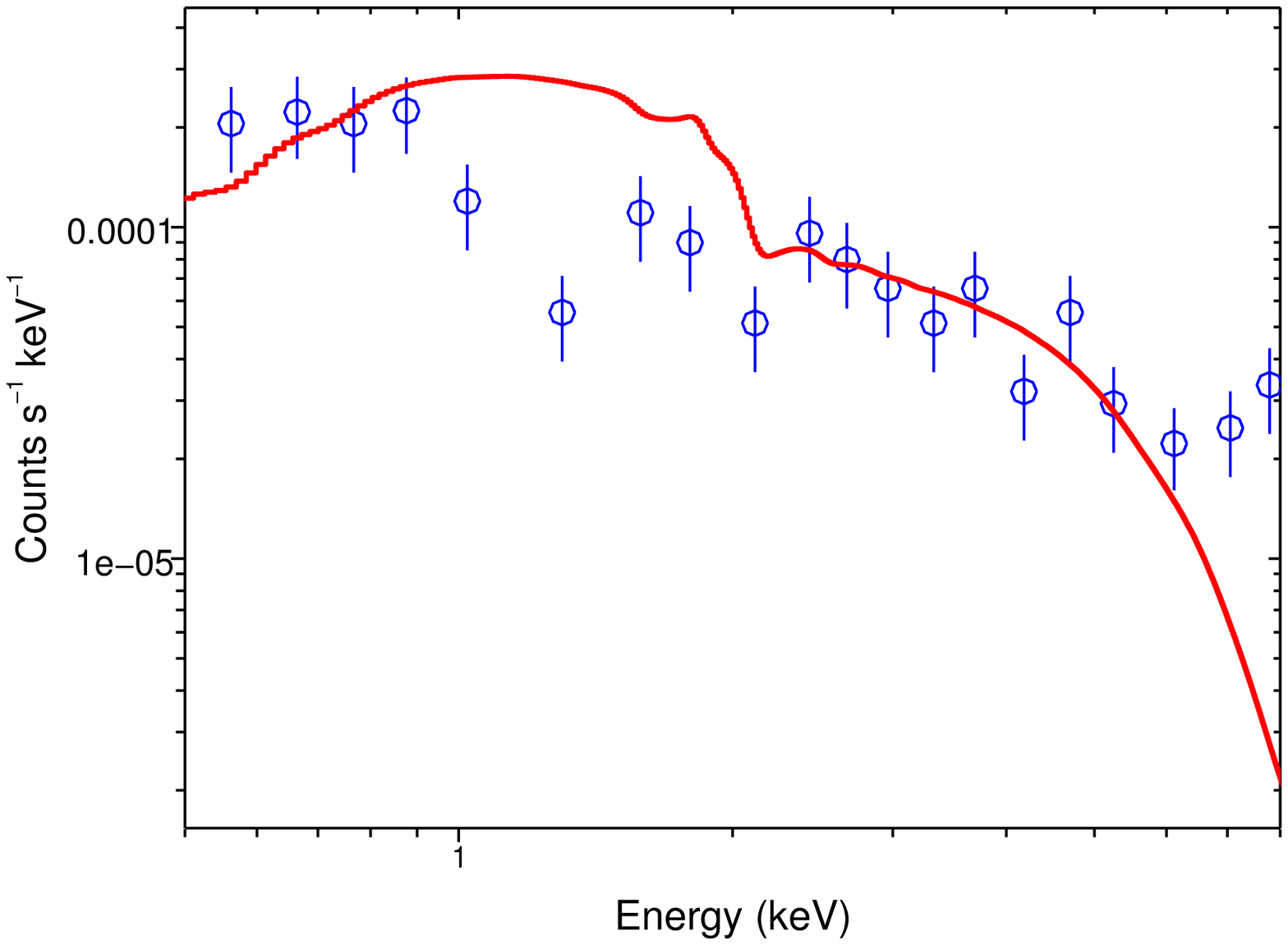}
    \caption{The spectrum and best fit powerlaw model extracted from the 2011 observations of SN 1957D. Note that while the spectrum shown here was grouped by 12 counts for ease of visibility, the model was produced by fitting the ungrouped spectrum.}
    \label{fig:sn1957d_spec_2}
\end{figure}

The supernova was clearly detected in 2010 and 2011 {\it Chandra}
observations, as previously reported in \citet{Long_2012}. The
spectrum from each individual dataset was extracted using a
4-arcsecond circular source region, centered on the source coordinates
(RA 13$^h$37$^m$03$^s$.584 DEC -29$^o$49'40.91'', from
\citet{Long_2012}), and a 4-arcsecond circular background region near
the source. The spectra were co-added using the \textsc{ciao}
\textit{combine\_spectra} script. The co-added, ungrouped source and
background spectra were simultaneously fit, using the \textit{cstat}
statistic. Several counts were present above 2 keV, and the spectra
were equally well fit using either a thermal or a non-thermal model -
an absorbed thermal plasma \textit{vmekal} model, and an absorbed
\textit{powerlaw} model, using \textit{xstbabs} model for the absorption
component with the column density frozen at the Galactic value. The
\textit{vmekal} fit gave a temperature of 3.44 keV for the 2010
dataset and 3.35 keV for the 2011 one. The powerlaw fit resulted in a
photon index of 1.06 for the 2010 observations and 1.07 for the 2011
observations.The spectrum from the 2011 observations, as well as the best fit thermal and powerlaw models, are shown in Figures \ref{fig:sn1957d_spec_1} and \ref{fig:sn1957d_spec_2}.

In 2014, the number of counts was too low to allow for fitting (a
4-arcsecond source region gave only 4 counts in the 0.3-8 keV
band). Instead the count rate at the source location was found using
the \textit{srcflux} command of the \textsc{ciao} software. The count
rate was converted to a flux value using the Chandra \textsc{pimms}
tool, with two cases taken into account: a plasma MEKAL model with
Galactic column density and a temperature of 3.43 keV; and a powerlaw
model with Galactic column density and photon index 1.06. These model
parameters were chosen based on the spectral fits from the 2010 and
2011 epochs.

The fluxes for the thermal and non-thermal model are presented in
Tables \ref{tab:sn1957d_1} and \ref{tab:sn1957d_2} respectively.

\begin{table}
    \centering
    \begin{tabular}{c c c c}
         \hline
         \colhead{Year} & \colhead{Observation ID} & \colhead{Date} & \colhead{Total Exposure (ks)} \\
         \hline
         2000 & 793 & 2000 Apr 29 & 51 \\
         2001 & 2064 & 2001 Sep 4 & 10 \\
         2010 & 12995 & 2010 Dec 23 & 59 \\
              & 13202 & 2010 Dec 25 & 99 \\
              & & & Total 158 \\
         2011 & 12993 & 2011 Mar 15 & 49 \\
              & 13241 & 2011 Mar 18 & 79 \\
              & 12994 & 2011 Mar 23 & 150 \\
              & 12996 & 2011 Mar 29 & 53.5 \\
              & 13248 & 2011 Apr 3 & 54.5 \\
              & 14332 & 2011 Aug 29 & 52.5 \\
              & 12992 & 2011 Sep 4 & 66.5 \\
              & 14342 & 2011 Dec 28 & 67 \\
              & & & Total 572 \\
         2014 & 16024 & 2014 Jun 7 & 30 \\
         \hline
    \end{tabular}
    \caption{Details of archival \textit{Chandra} observations of SN1957D}
    \label{tab:sn1957d_obs}
\end{table}

\begin{table}
    \centering
    \begin{tabular}{c c c c c c c c}
        \hline 
         \colhead{Instrument} & \colhead{Date of} & \colhead{Days since} & \colhead{Count rate} &  \colhead{kT (keV)} & \colhead{Reduced} & \colhead{Flux (0.3 - 2 keV)} & \colhead{Flux (0.3 - 8 keV)} \\
         \colhead{} & \colhead{observation} & \colhead{discovery} & \colhead{($\rm 10^{-4}~counts~s^{-1}$)} & \colhead{} & \colhead{statistic} & \colhead{($\rm 10^{-16}~erg~cm^{-2}~s^{-1}$)} & \colhead{($\rm 10^{-16}~erg~cm^{-2}~s^{-1}$)} \\
        \hline
         Chandra &  2000 Apr 29 & 15659 & $<$ 3.87 & 3.43 \footnote{\label{note2}Assumed value} & - & $<$ 12.69 & $<$ 23.37 \\
         (ACIS-S) & & & & & & \\
         Chandra & 2001 Sep 4 & 16152 & $<$ 10.4 & 3.43$^{\rm \ref{note2}}$ & - & $<$ 34.11 & $<$ 62.81 \\
         (ACIS-S) & & & & & & \\ 
         Chandra & 2010 Dec 23-25 & 19550 & 2.91 $\pm$ 1.07 & 3.44$_{-0.94}^{+75.5\footnote{\label{note3}No upper limit found}}$ & 0.42 & 5.29 $\pm$ 1.84 & 10.35 $\pm$ 3.19 \\
         (ACIS-S) & & & & & & \\
         Chandra & 2011 Mar 15 & 19775 & 2.34 $\pm$ 0.56 & 3.35$_{-0.46}^{+75.5^{\rm \ref{note3}}}$ & 1.05 & 4.05 $\pm$ 1.05 & 7.44 $\pm$ 1.97 \\
         (ACIS-S) & - Dec 28 & & & & & \\
         Chandra & 2014 Jun 7 & 20811 & 0.78 $\pm$ 0.73 & 3.43$^{\rm \ref{note2}}$ & - & 4.95 $\pm$ 4.63 & 9.12 $\pm$ 8.53 \\
         (ACIS-I) & & & & & & \\
        \hline
    \end{tabular}
    \caption{X-ray data for SN 1957D, assuming a plasma MEKAL
      model. Where possible, 1-$\sigma$ (68\%) error-bars are
      given. The column density is fixed to the Galactic value of
      3.94$\times$ 10$^{20}$ cm$^{-2}$ }
    \label{tab:sn1957d_1}
\end{table}

\begin{table}
    \centering
    \begin{tabular}{c c c c c c c c}
        \hline 
         \colhead{Instrument} & \colhead{Date of} & \colhead{Days since} & \colhead{Count rate} &  \colhead{Photon} & \colhead{Reduced} & \colhead{Flux (0.3 - 2 keV)} & \colhead{Flux (0.3 - 8 keV)} \\
         \colhead{} & \colhead{observation} & \colhead{discovery} & \colhead{($\rm 10^{-4}~counts~s^{-1}$)} & \colhead{Index} & \colhead{statistic} & \colhead{($\rm 10^{-15}~erg~cm^{-2}~s^{-1}$)} & \colhead{($\rm 10^{-15}~erg~cm^{-2}~s^{-1}$)} \\
        \hline
         Chandra & 2000 Apr 29 & 15659 & $<$ 3.87 & 1.06 \footnote{\label{note4}Assumed value} & - & $<$ 0.96 & $<$ 4.04 \\
         (ACIS-S) & & & & & & \\
         Chandra & 2001 Sep 4 & 16152 & $<$ 10.4 & 1.06$^{\rm \ref{note4}}$ & - & $<$ 2.57 & $<$ 10.87 \\
         (ACIS-S) & & & & & &\\
         Chandra & 2010 Dec 23-25 & 19550 & 2.91 $\pm$ 1.07 & 1.06 $\pm$ 0.39 & 0.42 & 1.81 $\pm$ 0.43 & 7.77 $\pm$ 5.28 \\
         (ACIS-S) & & & & & &\\
         Chandra & 2011 Mar 15 & 19775 & 2.34 $\pm$ 0.56 & 1.07$_{-0.17}^{+0.35}$ & 1.06 & 1.25 $\pm$ 0.16 & 5.27 $\pm$ 1.29 \\
         (ACIS-S) & - Dec 28 & & & & & \\
         Chandra & 2014 Jun 7 & 20811 & 0.78 $\pm$ 0.73 & 1.06$^{\rm \ref{note4}}$ & - & 0.30 $\pm$ 0.28 & 1.28 $\pm$ 1.20 \\
         (ACIS-I) & & & & & & \\
        \hline
    \end{tabular}
    \caption{X-ray data for SN 1957D, assuming a non-thermal powerlaw
      model. Where possible, 1-$\sigma$ (68\%) error-bars are
      given. The column density is fixed to the Galactic value of 3.94
      $\times$ 10$^{20}$ cm$^{-2}$}
    \label{tab:sn1957d_2}
\end{table}

\subsection{SN 1959D} 
\label{subsec:sn1959d}

SN 1959D lies in the galaxy NGC 7331. {This galaxy is viewed almost edge-on, hence the column density through the galaxy must also be taken into account. The SN} was first observed on 1959
June 28 \citep{1959d_discovery}, with a photographic magnitude of 13.4
\citep{1957d_discovery}. It was not detected in a radio search for
intermediate age SNe \citep{ecketal02}, but was detected in an X-ray
archival search \citep{Soria_2008}.

In this paper we consider 5 sets of \textit{Chandra} ACIS-S
observations to study the X-ray emission of SN 1959D,taken in 2001, 2014, 2015, 2016 and 2017 respectively. The details of these observations are given in Table \ref{tab:sn1959d_obs}. None of these observations had sufficient counts to allow for spectral fitting - using a 4-arcsecond source region, there were 7 counts in the 0.3-8 keV band in 2001, 3 in 2014, 4 in 2015, and 10 each in 2016 and 2017. For the 2001
observation, it was possible to use the command \textit{srcflux} to
extract a count rate at the source position. For all the other
observations, \textit{srcflux} gave either 0 counts or did not yield a
result, and could not be used even to find the upper limit on the
count rates. Therefore for these observations, the tables of confidence limits from \citet{kraft1991} were used to find the 99\% confidence level upper limits on the count rates. \citet{kraft1991} use Bayes theorem with a non-negative prior to derive the confidence limits on the source counts, given the number of observed counts in the vicinity of the source and background counts. Their method is applicable in cases where the number of counts is sufficiently low to be described by Poissonian statistics, and allows for the background counts to be greater than the observed counts. The fluxes were estimated from the count rates by using the \textit{Chandra} \textsc{pimms} tool, assuming a plasma MEKAL model with a temperature
of 0.61 keV (as for SN 1970G). {The value of the column density was  fixed at 2.0 $\times$ 10$^{21}$ cm$^{-2}$ \citep{Stockdale1998}, which takes into account the Galactic column density as well as that through the bulge of the host galaxy NGC 7331}. The
count rates and resultant fluxes are presented in Table
\ref{tab:sn1959d}.

\begin{table}
    \centering
    \begin{tabular}{c c c c}
        \hline
         \colhead{Year} & \colhead{Observation ID} & \colhead{Date} & \colhead{Total Exposure (ks)}  \\
         \hline
         2001 & 2198 & 2001 Jan 27 & 29 \\
         2014 & 16005 & 2014 Nov 3 & 10 \\
         2015 & 17569 & 2015 Jan 30 & 10 \\
              & 17570 & 2015 Apr 20 & 10 \\
              & 17571 & 2015 Aug 28 & 10 \\
              & & & Total 30 \\
         2016 & 18340 & 2016 May 5 & 28 \\
              & 18341 & 2016 Oct 25 & 30 \\
              & & & Total 58 \\
         2017 & 18342 & 2017 Jun 9 & 28 \\
         \hline 
    \end{tabular}
    \caption{Details of archival \textit{Chandra} observations of SN1959D}
    \label{tab:sn1959d_obs}
\end{table}

\begin{table}
    \centering
    \begin{tabular}{c c c c c c }
        \hline
        \colhead{Instrument} & \colhead{Date of} & \colhead{Days since} & \colhead{Count rate} & \colhead{Flux (0.3 - 2 keV)} & \colhead{Flux (0.3 - 8 keV)} \\
         \colhead{} & \colhead{observation} & \colhead{discovery} & \colhead{($\rm 10^{-4}~counts~s^{-1}$)} & \colhead{($\rm 10^{-15}~erg~cm^{-2}~s^{-1}$)} & \colhead{($\rm 10^{-15}~erg~cm^{-2}~s^{-1}$)} \\
         \hline 
         Chandra  & 2001 Jan 27 & 15189 & 1.59 $\pm$ 0.89 & 0.96 $\pm$ 0.54 & 0.99 $\pm$ 0.55 \\
         (ACIS-S) & & & & & \\
         Chandra  & 2014 Nov 3 & 20217 & $<$ 7.68 & $<$ 7.20 & $<$ 7.40 \\
         (ACIS-S) & & & & & \\
         Chandra  & 2015 Jan 30 - Aug 28 & 20410 & $<$ 2.28 & $<$ 2.14 & $<$ 2.19 \\
         (ACIS-S) & & & & & \\
         Chandra  & 2016 May 5 - Oct 24 & 20852 & $<$ 1.69 & $<$ 2.25 & $<$ 2.32 \\
         (ACIS-S) & & & & & \\
         Chandra  & 2017 Jun 9 & 21166 & $<$ 5.60 & $<$ 8.14 & $<$ 8.36 \\
         (ACIS-S) & & & & & \\
         \hline 
    \end{tabular}
    \caption{X-ray data for SN 1959D, assuming a plasma MEKAL
      model. The temperature is chosen to be 0.61 keV, and the column
      density is fixed to the value of 2.0 $\times$
      10$^{21}$ cm$^{-2}$, {which takes into account the column density along the line-of-sight towards NGC 7331 as well as that through the bulge of the host galaxy NGC 7331}.}
    \label{tab:sn1959d}
\end{table}

\subsection{SN 1941C} 
\label{subsec:sn1941c}

\begin{figure}
    \centering
    \includegraphics[width=0.7\linewidth]{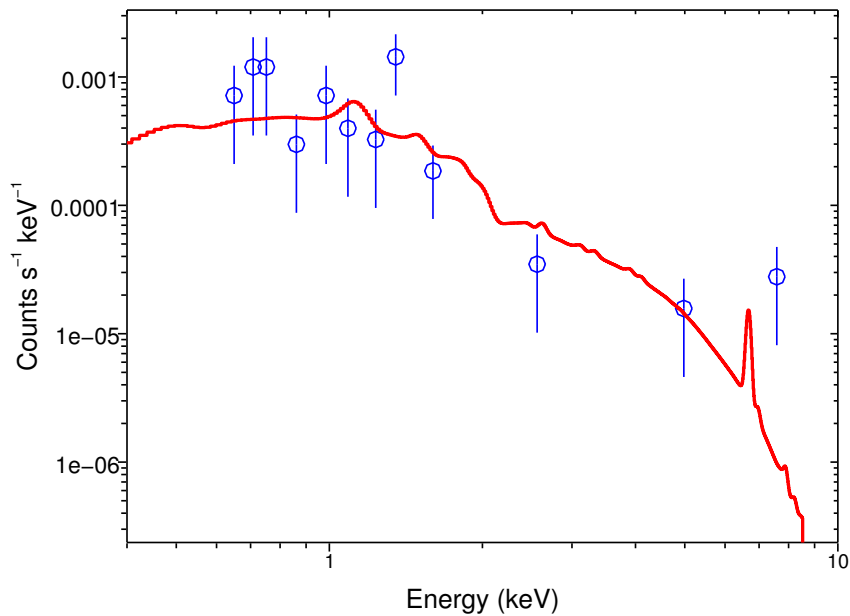}
    \caption{The spectrum and best fit model extracted from the 2002 observation of SN 1941C. Note that while the spectrum shown here was grouped by 2 counts for ease of visibility, the model was produced by fitting the ungrouped spectrum}
    \label{fig:sn1941c_spec}
\end{figure}

SN 1941C is probably the oldest SN to be detected in X-rays
\citep{Soria_2008}. It was discovered on 1941 April 16
\citep{1941c_discovery}.  Only a single set of \textit{Chandra}
observations is available for SN 1941C, taken between 2002 March 7 and
2002 June 8 (ObsID 2920, 2921) using the ACIS-S instrument, for a total exposure time of 38 ks. The source and background spectra were
extracted using a 4-arcsecond circular source region and a 4-arcsecond
circular background region, arbitrarily chosen near the source (source
coordinates from the SIMBAD astronomical database). The spectra were
co-added using \textsc{ciao} \textit{combine\_spectra} script. The
co-added source and background spectra were simultaneously fit, using
the \textit{cstat} statistic. An absorbed thermal plasma
\textit{vmekal} model, with \textit{xstbabs} as the absorption
component, was used. The column density was fixed at the Galactic
value of 1.64 $\times$ 10$^{20}$ cm$^{-2}$ \citep{DL_nH}. Spectral
fitting resulted in a temperature of 3.70$^{+2.40}_{-1.39}$ keV, with a  reduced statistic of 0.17. The resultant fluxes were 2.28 $\pm$ 0.55 $\times$ 10$^{-15}$ ergs
cm$^{-2}$ s$^{-1}$ in the 0.3 - 2 keV band and 4.34 $\pm$ 1.12
$\times$ 10$^{-15}$ ergs cm$^{-2}$ s$^{-1}$ in the 0.3 - 8 keV band. The spectrum and best fit model are shown in Figure \ref{fig:sn1941c_spec}.

\section{Results} 
\label{sec:results}

\begin{figure}
    \centering
    \includegraphics[width=\linewidth]{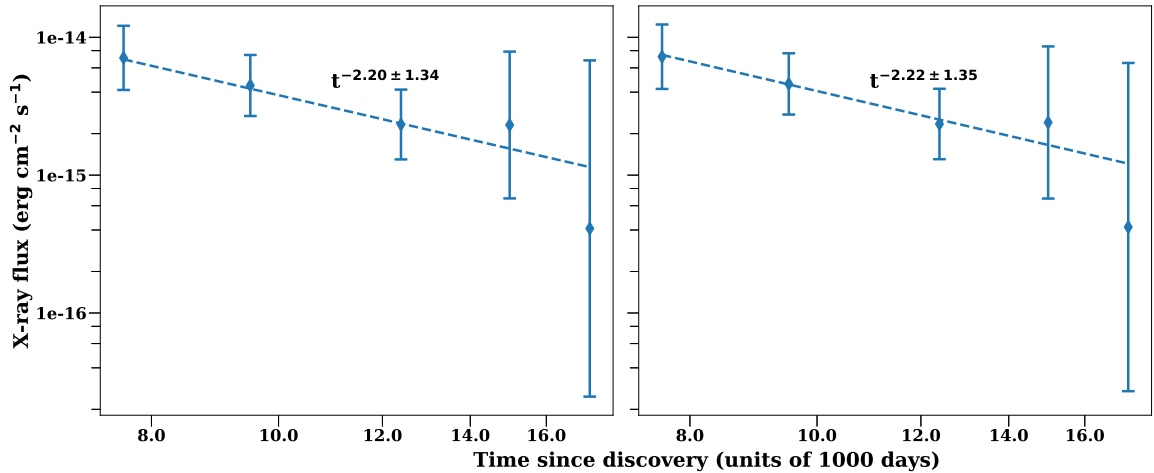}
    \caption{{(Left)} The 0.3 - 2 keV X-ray light curve for
      SN1970G. {(Right)} The 0.3 - 8 keV X-ray light curve for
      SN1970G.  In both cases we plot unabsorbed fluxes with
      1-$\sigma$ error bars. We fit the light curves with a powerlaw
      decrease with time; the best fits are shown as dashed lines. The
      best fitting models had flux $\propto$ t$^{-2.20 \pm 1.34}$, for
      the 0.3 - 2 keV light curve, and flux $\propto$ t$^{-2.22 \pm
        1.35}$, for the 0.3 - 8 keV light curve.}
    \label{fig:sn1970g}
\end{figure}

\begin{figure}
    \centering
    \includegraphics[width=\linewidth]{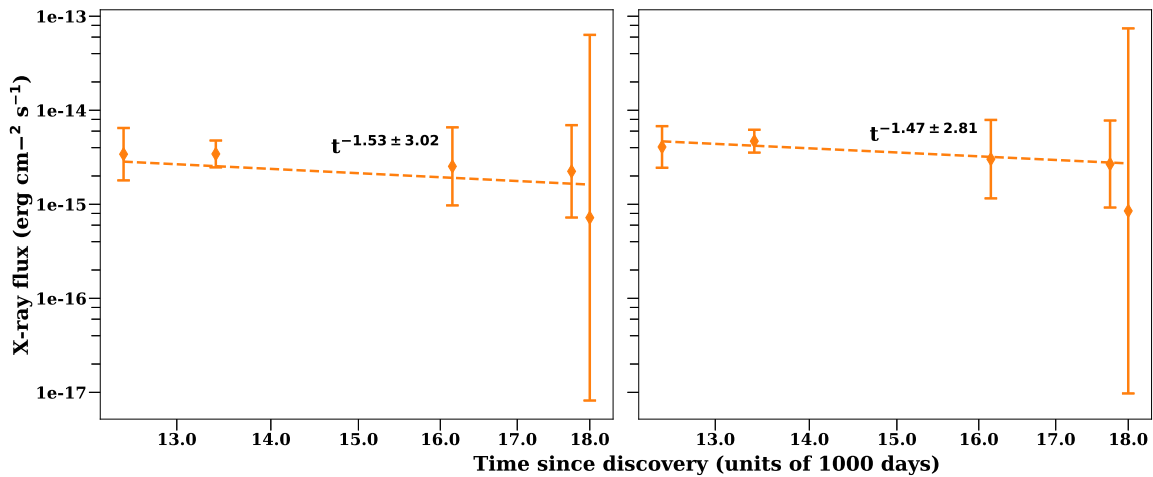}
    \caption{{(Left)} The 0.3 - 2 keV X-ray light curve for
      SN 1968D. {(Right)} The 0.3 - 8 keV X-ray light curve for
      SN 1968D. Other details similar to Figure \ref{fig:sn1970g}. The
      best fitting models were flux $\propto$ t$^{-1.53 \pm 3.02}$,
      for the 0.3 - 2 keV light curve, and flux $\propto$ t$^{-1.47
        \pm 2.81}$, for the 0.3 - 8 keV light curve.}
    \label{fig:sn1968d}
\end{figure}

\begin{figure}
    \centering
    \includegraphics[width=\linewidth]{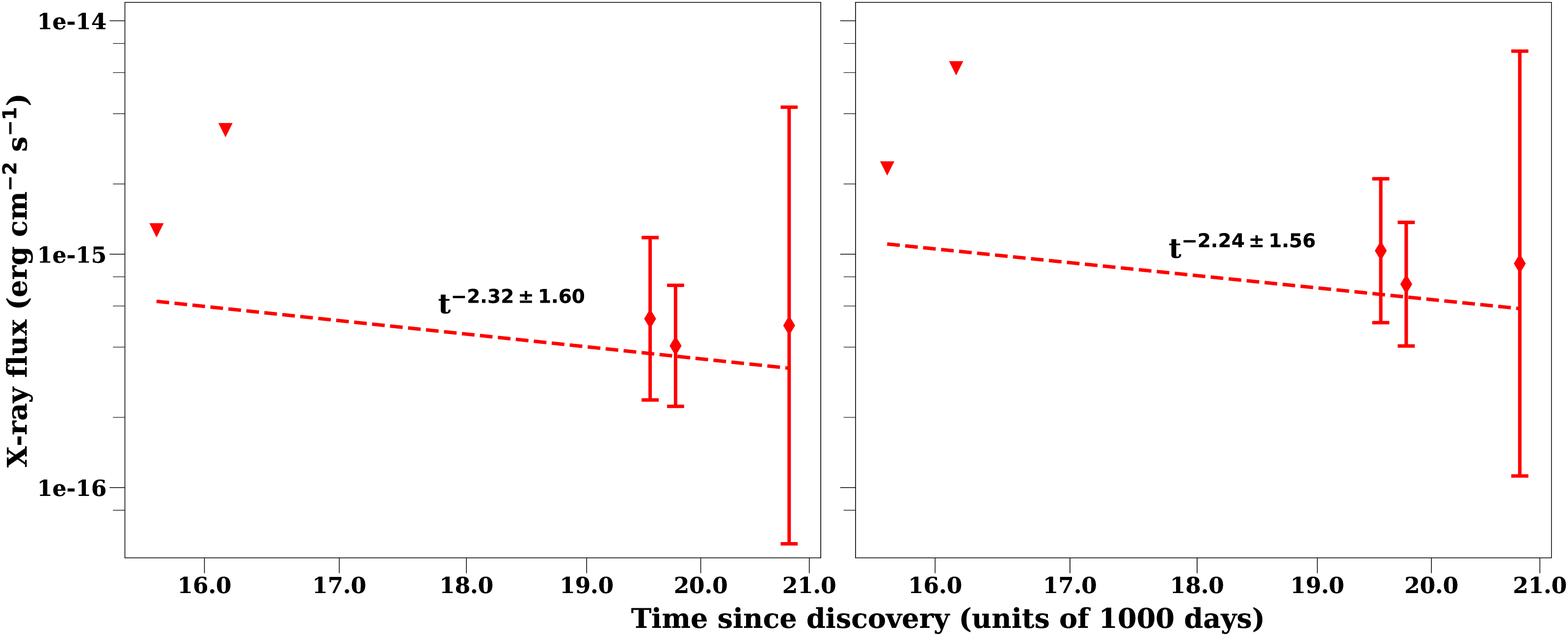}
    \caption{{(Left)} The 0.3 - 2 keV X-ray light curve for
      SN 1957D. {(Right)} The 0.3 - 8 keV X-ray light curve for
      SN 1957D. The unabsorbed fluxes were found assuming a thermal plasma model. Triangles indicate upper limits.  Other details similar to Figure \ref{fig:sn1970g}.  The best fitting models were
      flux $\propto$ t$^{-2.32 \pm 1.60}$, for the 0.3 - 2 keV light
      curve, and flux $\propto$ t$^{-2.24 \pm 1.56}$, for the 0.3 - 8
      keV light curve.}
    \label{fig:sn1957d_1}
\end{figure}

\begin{figure}
    \centering
    \includegraphics[width=\linewidth]{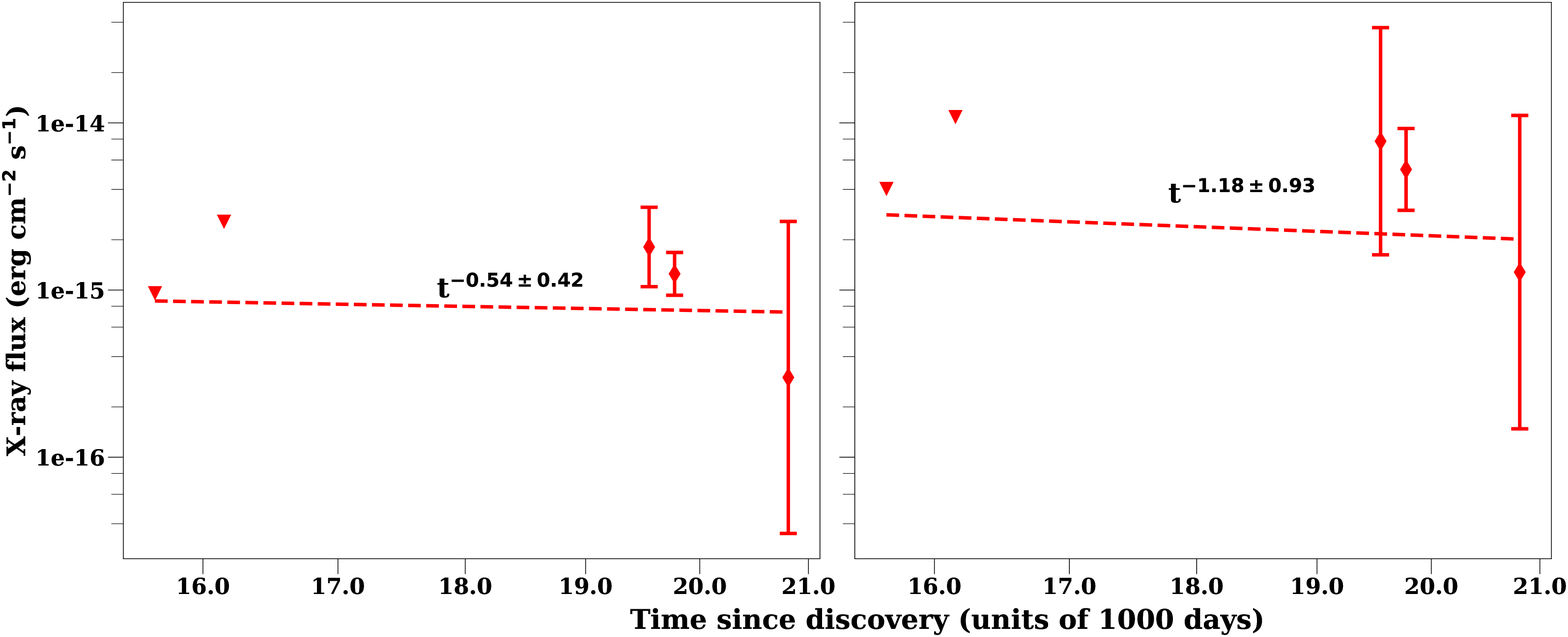}
    \caption{{(Left)} The 0.3 - 2 keV X-ray light curve for
      SN 1957D. {(Right)} The 0.3 - 8 keV X-ray light curve for
      SN 1957D. The unabsorbed fluxes were found assuming a power law model. Triangles indicate upper limits. Other details similar
      to Figure \ref{fig:sn1970g}. The best fitting models were flux
      $\propto$ t$^{-0.54 \pm 0.42}$, for the 0.3 - 2 keV light curve,
      and flux $\propto$ t$^{-1.18 \pm 0.93}$, for the 0.3 - 8 keV
      light curve.}
    \label{fig:sn1957d_2}
\end{figure}

The X-ray light curves  of SN 1970G, SN 1968D and SN 1957D are presented in Figs \ref{fig:sn1970g}, \ref{fig:sn1968d},
\ref{fig:sn1957d_1} and \ref{fig:sn1957d_2}. All three supernovae
display a {tentative} decrease in flux with time, {albeit with large error bars}.  Based on analytical
models for the circumstellar interaction and X-ray emission (see for
example \citet{Fransson1996, Dwarkadas_2012}), we expect the flux to
have a power-law decrease with time. Therefore the light curves were
fit with a model of the form L$_X \propto$ t$^{-\alpha}$. In each case, the expectation values and 1-$\sigma$ errors were found by sampling the posterior probability distribution of $\alpha$ using the Markov Chain Monte Carlo method. Table \ref{tab:slopes} gives the best fit values of
$\alpha$ for the 0.3 - 2 keV and 0.3 - 8 keV band light curves. It is clear that there is considerable uncertainty in the slopes, due to the progressively larger errorbars on the flux values as the supernovae became older and fainter. As the most recent observations, in all cases, had insufficient counts to allow for spectral fitting, a more conservative estimate of the slopes of the light curves was also made, with the {\it srcflux} command used to derive only an upper limit instead of a flux value for these observations. These conservative slopes, given in Table \ref{tab:slopes_2}, are somewhat steeper.
\begin{table}
    \centering
    \begin{tabular}{c c c}
        \hline
        \colhead{Supernova} & \colhead{$\alpha$ - 0.3 - 2 keV band} & \colhead{$\alpha$ - 0.3 - 8 keV band}\\
        \hline 
         SN 1970G & 2.20 $\pm$ 1.34 & 2.22 $\pm$ 1.35 \\  
         SN 1968D & 1.53 $\pm$ 3.02 & 1.47 $\pm$ 2.81 \\ 
         SN 1957D (\textit{vmekal} model) & 2.32 $\pm$ 1.60 & 2.24 $\pm$ 1.56 \\
         SN 1957D (\textit{powerlaw} model) & 0.54 $\pm$ 0.42 & 1.18 $\pm$ 0.93 \\
        \hline 
    \end{tabular}
    \caption{The best fit powerlaw index $\alpha$ for the lightcurves of SN 1970G, SN 1968D and SN 1957D}
    \label{tab:slopes}
\end{table}

\begin{table}
    \centering
    \begin{tabular}{c c c}
        \hline
        \colhead{Supernova} & \colhead{$\alpha$ - 0.3 - 2 keV band} & \colhead{$\alpha$ - 0.3 - 8 keV band}\\
        \hline 
         SN 1970G & 2.58 $\pm$ 1.11 & 2.58 $\pm$ 1.12 \\  
         SN 1968D & 2.06 $\pm$ 2.47 & 1.80 $\pm$ 2.56 \\ 
         SN 1957D (\textit{vmekal} model) & 2.33 $\pm$ 1.62 & 2.23 $\pm$ 1.54 \\
         SN 1957D (\textit{powerlaw} model) & 0.54 $\pm$ 0.42 & 1.17 $\pm$ 0.92 \\
        \hline 
    \end{tabular}
    \caption{The best fit powerlaw index $\alpha$ for the lightcurves of SN 1970G, SN 1968D and SN 1957D, with the most recent observation in each case used to determine only an upper limit and not a definite flux value. }
    \label{tab:slopes_2}
\end{table}

\begin{figure}
    \centering
    \includegraphics[width=\linewidth]{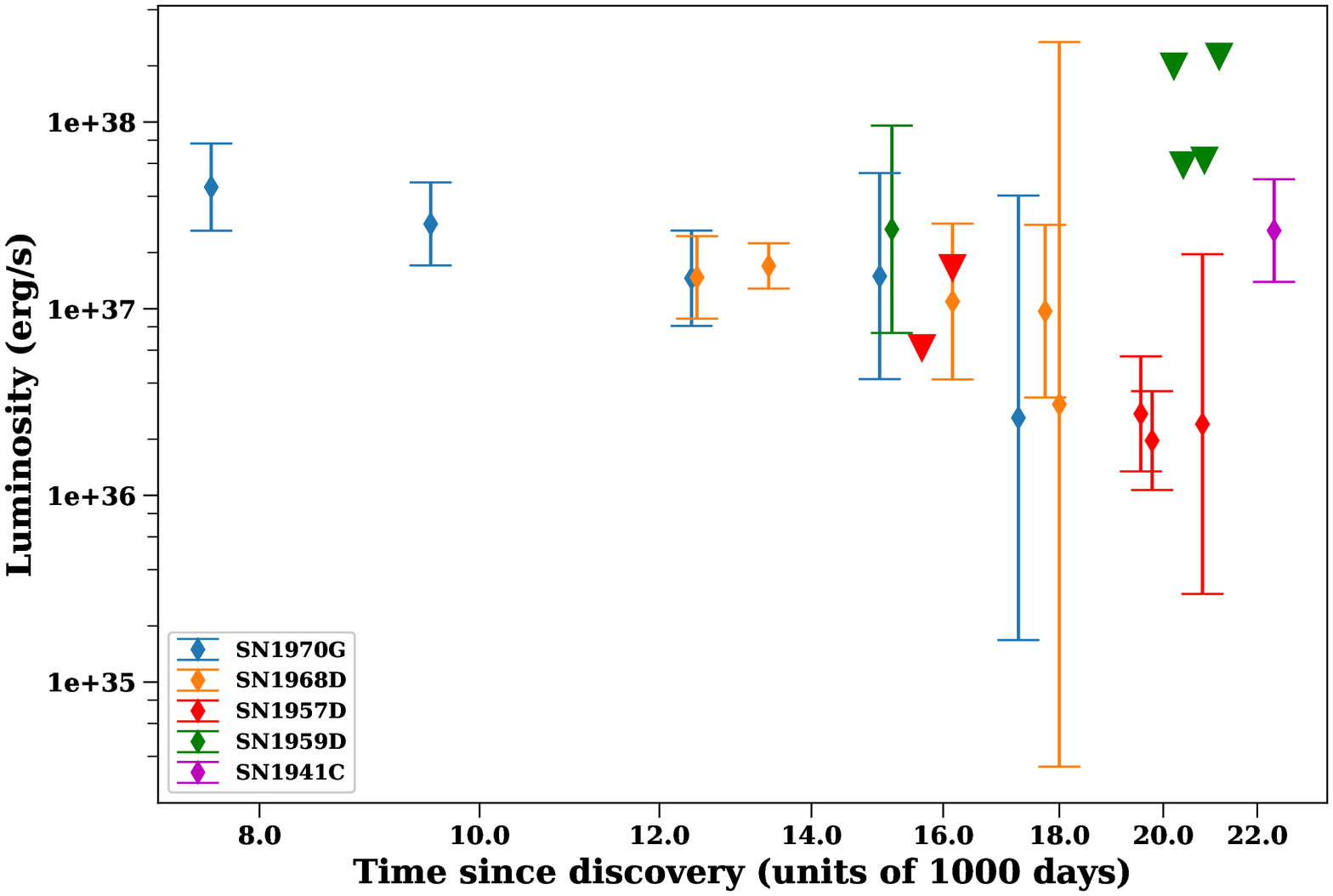}
    \caption{0.3 - 8 keV X-ray luminosities, plotted versus time, for
      all five supernovae studied in this paper. The axes are
      logarithmic. The different colours correspond to different
      supernovae, as shown in the legend. Where possible, we have
      given luminosities with 1-$\sigma$ error bars; triangles indicate
      upper limits. The distances used are given in the text.}
    \label{fig:all_sn}
\end{figure}

The existence of only a single Chandra observation for SN 1941C precludes a
lightcurve. It was not possible to construct an X-ray light curve for
SN 1959D, but upper limits on the flux at 20217, 20410, 20852 and 21166
days since the discovery of the SN were found. Figure \ref{fig:all_sn} shows the
luminosities and upper limits for all five supernovae. A distance of
7.2 Mpc for SN 1970G \citep{1970G}, 5.5 Mpc for SN 1968D \citep{1968D},
4.7 Mpc for SN 1957D \citep{1957D}, 9.7 Mpc for SN 1941C \citep{1941C},
and 15 Mpc for SN 1959D \citep{1959D} was adopted. All five supernovae
were found to have comparable luminosities (within two orders of magnitude) at the late epoch studied
here. This can be compared to overall luminosities of SNe, which span almost 9 orders of magnitude \citep{Dwarkadas_2012}.

For those SNe whose electron temperatures could be derived from
spectral fits, we can infer a shock velocity. From the
Rankine-Hugoniot shock jump conditions:

\begin{eqnarray}
T_e &=& \frac{3}{16}\frac{\mu m_p v_{sh}^2}{k}
\nonumber \\
v_{sh} &=& \left(\frac{16}{3}\frac{kT_e}{\mu m_p}\right)^{1/2}
\end{eqnarray}

Here T$_e$ is the electron temperature, m$_p$ the mass of a proton,
v$_{sh}$ the velocity of the shock, k the Boltzmann constant and $\mu$
the average atomic mass of the material in the shock. This equation is strictly valid only when the electron temperature is equal to the proton temperature, which is not necessarily true in SN shocks, which are collisionless shocks. The energy is preferably transferred to the ions, and the electron temperature is initially lower. Over time the electron to proton temperature ratio approaches 1, due to Coulomb collisions or various plasma processes. Since X-rays measure the electron temperature, the velocity so derived can be considered as a lower bound. For fully ionized solar plasma, $\mu$ is about 0.63, whereas for a neutral
plasma its value is about 1.3. At the temperatures derived here the
plasma will be ionized to some extent. For simplicity we take $\mu$ =
1, and find an average shock velocity $v_{sh}$ of 520 km s$^{-1}$ for
SN 1970G, 850 km s$^{-1}$ for SN 1968D, 1320 km s$^{-1}$ for SN 1957D and
1380 km s$^{-1}$ for SN 1941C. For a different value of $\mu$, the
velocity goes as $v_{sh} \times {\mu}^{-1/2}$.

\section{Discussion}

\subsection{Analysis of individual supernovae} \label{subsec:individual}

{\textit{SN 1970G:}} Assuming that the SN 1970G lightcurve is
decreasing as a power-law with time, we find that the best fit
powerlaw index to our data is 2.20 $\pm$ 1.34 for the 0.3 - 2 keV band
light curve (Table \ref{tab:slopes}). This is consistent with the best
fit powerlaw (2.7 $\pm$ 0.9 for the 0.3 - 2 keV band light curve)
reported in \citet{Immler_2005}. Our 2004 \textit{Chandra} flux is
similarly consistent with theirs, which is no surprise given that the
fitted temperature is close to the temperature that they assumed.\\

\begin{figure}
    \centering
    \includegraphics[width=0.7\linewidth]{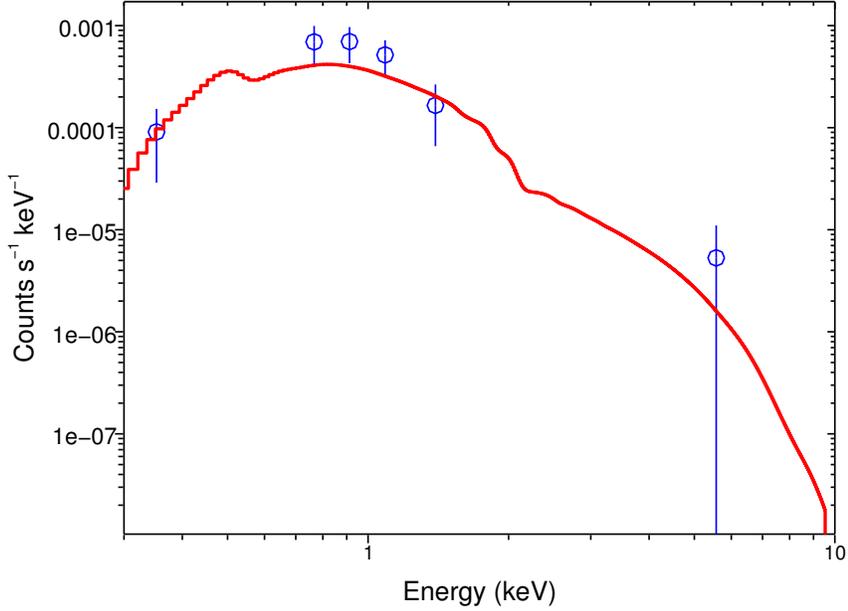}
    \caption{The best fit model powerlaw model for the 2004 observation of SN 1970G. The spectrum was grouped by 15 counts and fitted after background subtraction, with the absorbing column fixed at the Galactic value.}
    \label{fig:sn1970g_spec_pwlaw}
\end{figure}

\citet{midlife_crisis} analyzed the 2011 observation of the supernova,
and found that the X-ray flux had increased by a factor $\sim$ 3
compared to that in 2004. This was attributed to the switching-on of a
pulsar wind nebula by the authors. Our results are inconsistent with
this; while a possible increase in the flux from 2004 to 2011 could be
accommodated, the error bars on the flux in 2011 show that it is also
consistent with a steadily decreasing light curve, as shown in Figure
\ref{fig:sn1970g}. We suspect that the discrepancy arises because
\citet{midlife_crisis} modeled the source spectrum from the 2011
observation of SN 1970G with a powerlaw, i.e. assuming the emission
mechanism is non-thermal, but then compared them to the fluxes at the
previous epochs which were computed by \citet{Immler_2005} using a
thermal model. When we modeled the source spectra from the 2011 and
2017 observations using a powerlaw model, we also find the flux in
2011 to be much greater than that given by the thermal model in
2004. However, the lack of high-energy counts in both the 2011 and
2017 spectra are generally not indicative of a power-law, and there is
no other indication for the spectral shape and emission mechanism to
change at the 2011 epoch from a thermal to a non-thermal one. The 2004
spectrum is better fit with a thermal model (see Figures \ref{fig:sn1970g_spec_1} and \ref{fig:sn1970g_spec_pwlaw}), and we believe that the remaining spectra are better modelled as thermal plasmas
also (see Figure \ref{fig:sn1970g_spec_2}). The low count rate in the 2017 Chandra observation is not
consistent with a steady or increasing flux, as might be expected from
a new pulsar wind nebula.

\begin{figure}
    \centering
    \includegraphics[width=\linewidth]{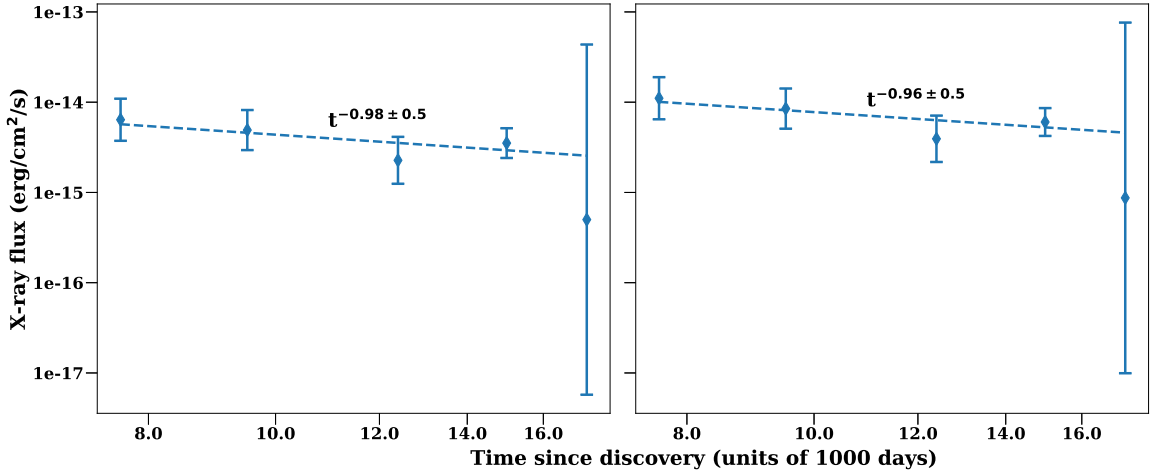}
    \caption{{(Left)} The 0.3 - 2 keV X-ray light curve for
      SN 1970G, with flux found using a powerlaw
      model. {(Right)} The 0.3 - 8 keV X-ray light curve for
      SN 1970G, with flux found using a powerlaw model. The light
      curves are plotted on logarithmic axes, with 1-$\sigma$ error
      bars. Though it is possible that the flux increased slightly
      from 2004 to 2011, the derived flux values are also consistent
      with a steady powerlaw decrease with time (as shown by the
      dotted line, the best fit model is t$^{-0.98 \pm 0.5}$).}
    \label{fig:sn1970g_plaw}
\end{figure}

Finally, we note that if all the available X-ray observations of
SN 1970G are modeled using powerlaws, then the values are again
consistent (within the errorbars) with a steady decrease of flux with
time (see Figure \ref{fig:sn1970g_plaw}), though the absolute value of
the flux at each epoch is higher than that found using a thermal
model. Therefore, using a thermal or non-thermal model uniformly over
all epochs does not result in an increasing flux during the
evolution. It is only the switch from a thermal to a non-thermal model
midway through the evolution that results in this increase, and we do
not find sufficient evidence pointing to a change in the emission
mechanism. Thus our results disfavor a model of a new pulsar wind
nebula in SN 1970G. 
\newline

{\textit{SN 1968D:}} The discovery of SN 1968D in X-rays was
first reported by \citet[][hereafter SP08]{Soria_2008}.  SP08 coadded
the spectra from the 2001, 2002 and 2004 observations.For a thermal plasma model, SP08 derived an unabsorbed luminosity in the 0.3 - 8 keV band of 15.1$^{+9.0}_{-8.1}$ $\times$ 10$^{37}$ erg s$^{-1}$. By comparison, we derive a luminosity of 1.47 $\pm$ 0.33 $\times$ 10$^{37}$ erg s$^{-1}$ for the 2001-02 observations and 1.69 $\pm$ 0.21 $\times$ 10$^{37}$ erg s$^{-1}$ for the 2004 observations, an order of magnitude lower. However, SP08 arrived at a temperature of 0.69 keV and an extremely high column density of 1.07 $\times$ 10$^{22}$ cm$^{-2}$. These results are inconsistent with ours; we found that all of our spectra were well-fit by a model with temperature $\sim$ 1.50 keV and Galactic column density (2.11 $\times$ 10$^{21}$ cm$^{-2}$). The large difference between the luminosities found in this work and in SP08 can likely be attributed to the large difference between the adopted values of column density.  Given the limited number of counts in the X-ray spectrum of such an old supernova, it is difficult to discuss the validity of various models. On fitting the 2001-02 and 2004 spectra with the temperature fixed at the value found in SP08 (0.69 keV), we found a similarly high column density of 1.19 $\times$ 10$^{22}$ cm$^{-2}$. The two models have the same reduced statistic, and so are statistically indistinguishable. On the other hand, the high column density would imply a large plasma density around the SN at an age $> 30$ yr. Unless there is a good reason to suspect the existence of a large amount of material ahead of the SN shock, a model that returns the Galactic column is preferable as discussed in the beginning of Section \ref{sec:analysis}. In other SNe, it is rare to see such high column densities even at an age of 10-20 years, let alone at the age of 33 years as for SN1968D.
\newline

{\textit{SN 1957D:}} The detection of SN 1957D in X-rays was
first reported in \citet[][hereafter L12]{Long_2012}, as part of an
extensive campaign to study M83 in X-rays \citep{Long_2014}. Our flux
values are in agreement with theirs, but the other fitted parameters
are not. L12 co-added the \textit{Chandra} observations of SN 1957D
taken in 2010 and 2011. In our study we found there were enough counts
at each epoch to analyze the 2010 and 2011 observations
separately. L12 were unable to distinguish between a thermal or a
non-thermal (powerlaw) model fit to the spectrum, consistent with our
modelling. They reported a photon index for the power-law model of
1.4$^{+1.0}_{-0.8}$, but could not constrain the temperature for the
\textit{vmekal} model, giving a lower limit of 10 keV. For both
models, they inferred a high column density of 2.0 $\times$ 10$^{22}$
cm$^{-2}$. The powerlaw photon index which they derived is consistent
within errorbars with our values (1.06 $\pm$ 0.39 and
1.07$^{+0.35}_{-0.17}$ for 2010 and 2011 respectively, see Table
\ref{tab:sn1957d_2}). The other results are quite different. We find a temperature of $\sim$ 3.40 keV for the thermal model, and our
spectra are well fit by models which fix the column density at the
Galactic value. L12 found a 0.3 - 8 keV luminosity of
1.7$^{+2.4}_{-0.3}$ $\times$ 10$^{37}$ erg s$^{-1}$ for the powerlaw
model. We find a 0.3 - 8 keV luminosity of 2.0 $\pm$ 1.4 $\times$
10$^{37}$ erg s$^{-1}$ and 1.4 $\pm$ 0.3 $\times$ 10$^{37}$ erg
s$^{-1}$ for the 2010 and 2011 observations respectively. The average value would be consistent with that from L12. L12 do not specify their fitting procedure in detail; therefore it is not possible to compare our method with theirs. As discussed earlier, we feel that for older SNe it is more appropriate to use a fit which assumes the Galactic column density, unless otherwise dictated by the observations.
\newline

{\textit{SN 1959D:}} A tentative detection (at a 90\% confidence level) of SN 1959D in X-rays was first reported in SP08 for the 2001 {\it Chandra} observation. The flux value which they derived is in good agreement with ours. The authors assumed a powerlaw model with powerlaw index 2 and a column density of 6.2 $\times$ 10$^{20}$ cm$^{-2}$ for the source, and reported an approximate luminosity of 3.5$\times$ 10$^{37}$ erg s$^{-1}$ in the 0.3 - 8 keV band, corresponding to a flux of 1.28 $\times$ 10$^{-15}$ erg cm$^{-2}$ s$^{-1}$. They found that the luminosity was constrained between 1 - 7 $\times$ 10$^{37}$ erg s$^{-1}$ with 90\% confidence level, corresponding to a flux interval of 0.40 - 2.56 $\times$ 10$^{-15}$ erg cm$^{-2}$ s$^{-1}$. This flux interval is in agreement with our value of 0.68 $\pm$ 0.38 $\times$ 10$^{-15}$ erg cm$^{-2}$ s$^{-1}$, although we made use of a thermal plasma model to describe the source, {and considered a higher column density including the column through the bulge of NGC 7331} (as discussed in Section \ref{subsec:sn1959d}).
\newline 

{\textit{SN 1941C:}} The discovery of SN 1941C in X-rays was
first reported by SP08, who found that the source could be fit with a
thermal plasma model, with temperature $\sim$ 3 keV. They derived a
flux in the 0.3 - 8 keV band of 4.40$^{+2.3}_{-1.7}$ $\times$
10$^{-15}$ erg cm$^{-2}$ s$^{-1}$. This is consistent with our results
within the error bars - we derived a temperature of 3.70 keV from
spectral fitting, and a flux in the 0.3 - 8 keV band of 4.34 $\pm$
1.12 $\times$ 10$^{-15}$ erg cm$^{-2}$ s$^{-1}$ s (as discussed in
Section \ref{subsec:sn1941c}).

\subsection{Analysis of light curves} \label{subsec:light_curves}

The thermal X-ray emission from a supernova depends on the electron
density of the medium into which the shock is evolving (n$_e$), the
cooling function of the medium ($\Lambda$), and the volume of the
emitting region (V) \citep[see for example][]{Dwarkadas_2012}:
\begin{eqnarray}
    L_X = \Lambda n_e^2V
\label{eqn:xray}
\end{eqnarray}
In the self-similar solution \citep{Chevalier1982}, both the ejecta
density $\rho_{ej}$ and the CSM density $\rho_{CSM}$ have powerlaw
profiles. The density profiles for the ejecta and CSM are given by,
\begin{eqnarray} \label{eqn:rho_profile}
    \rho_{ej} \propto \left(\frac{r}{t}\right)^{-n}t^{-3}
    \nonumber \\
    \rho_{CSM} \propto r^{-s}
\end{eqnarray}
If n $>$ 5 and s $<$ 3, the evolution of the supernova is described by
the self-similar driven wave solution. The X-ray emission is assumed
to come from a thin shell of shocked material.  The radius of the
forward and reverse shock is proportional to the radius of the contact
discontinuity, which is given by,
\begin{eqnarray} \label{eqn:self_sim_r}
    r_{c} \propto t^{(n-3)/(n-s)}
\end{eqnarray}
Given the ages of the SNe we are studying, the X-ray emission could
conceivably originate from either the forward shock or the reverse
shock. From equation (\ref{eqn:self_sim_r}) we have for the density of
the material at the forward and reverse shocks,
\begin{eqnarray}
    \rho_{ej} &\propto& r^{-n}t^{n-3} \propto t^{-n(n-3)/(n-s)}t^{(n-3)}
    \nonumber \\
    \rho_{ej} &\propto& t^{-s(n-3)/(n-s)}
    \nonumber \\
    \rho_{CSM} &\propto& r^{-s}
    \nonumber \\
    \rho_{CSM} &\propto& t^{-s(n-3)/(n-s)}
\end{eqnarray}
Thus the density of the shocked material has the same time dependence
for both the forward and reverse shock, as expected for a self-similar
evolution. In the self-similar case the thickness of the shell will be
proportional to its radius, $\Delta r \propto r$, and so the shocked
volume depends on radius as V $\propto$ r$^3$. For electron
temperatures T$_e$ $\leq$ 2.6 $\times$ 10$^7$ K, or 2.24 keV, the
cooling function has the form $\Lambda \propto T_e^{-0.48}$, and for
T$_e$ $>$ 2.6 $\times$ 10$^7$ K $\Lambda \propto T_e^{0.5}$
\citep{Chevalier2017}. In either case, for a strong shock the
temperature at the shock depends on the shock velocity as T$_e$
$\propto$ v$_{sh}^2$ and v$_{sh}$ $\propto$ r$_{sh}$/t. As a result,
for T$_e$ $>$ 2.6 $\times$ 10$^7$ K, $\Lambda \propto$ r$_{sh}/t$, and
for T$_e$ $\leq$ 2.6 $\times$ 10$^7$ K, $\Lambda \propto$
(r$_{sh}/t)^{-0.96}$.  Thus for T$_e$ $>$ 2.6 $\times$ 10$^7$ K, the
X-ray luminosity evolves as,
\begin{eqnarray}\label{eqn:T_gtr_L}
    L_X &\propto& \frac{r_{sh}}{t}t^{-2s(n-3)/(n-s)}r_{sh}^3 = r_{sh}^{4}t^{-1-2s(n-3)/(n-s)}
    \nonumber \\
    L_X &\propto& t^{-[(2s-3)n + 12 - 7s]/(n-s)}
\end{eqnarray}
And for T$_e$ $\leq$ 2.6 $\times$ 10$^7$ K, the X-ray luminosity
evolves as,
\begin{eqnarray}\label{eqn:T_less_L}
    L_X &\propto& \left(\frac{r_{sh}}{t}\right)^{-0.96}t^{-2s(n-3)/(n-s)}r_{sh}^3 = r_{sh}^{2.04}t^{0.96-2s(n-3)/(n-s)}
    \nonumber \\
    L_X &\propto& t^{-[(2s-3)n + 6.12 - 5.04s]/(n-s)}
\end{eqnarray}
This is an equation for the X-ray luminosity over the entire X-ray
range. In general \textit{Chandra} detects X-ray emission in a narrow
energy band (0.3 - 10 keV). However, most of the X-ray emission from
the older SNe studied here appears to peak at low energies, as
evidenced by the low electron temperatures ($<$ 5 keV) obtained from
spectral fitting. Thus it seems reasonable to assume that the 0.3-8
keV band generally includes all the X-ray emission, and therefore the
decline of the 0.3 - 8 keV band X-ray luminosity can be approximated
using the above equations. \\
If the mass-loss parameters are constant with time, the wind is referred to as a steady wind. For such a wind, s = 2, and the equations \ref{eqn:T_gtr_L} and
\ref{eqn:T_less_L} become,
\begin{eqnarray}
    L_X = \left\{
        \begin{array}{ll}
            t^{-1} & \quad T_e > 2.6~\times~10^7~K \\
            t^{-(n-3.96)/(n-2)} & \quad T_e \leq 2.6~\times~10^7~K 
        \end{array}
    \right.
\end{eqnarray}

{We can compare these expressions to the power-law indices derived earlier (Tables \ref{tab:slopes} and \ref{tab:slopes_2}) The luminosity of SN 1970G is clearly decreasing with time. The best fit value of the powerlaw decrease would suggest a non-steady wind, i.e wind parameters varying with time. However the large errorbars complicate the interpretation - the slope is consistent within 1$\sigma$ with expansion into a steady wind. A steady wind would imply $n \ga $ 16, which is larger than the value  $9 < n <11$ generally characteristic of stellar envelopes \citep{mm99}.}

{The best fit powerlaw indices given in Table \ref{tab:slopes} and \ref{tab:slopes_2} suggest a decreasing lightcurve for SN 1968D, but the errobars are large enough that the X-ray luminosity could equally well be constant with time. A constant luminosity gives $s \sim$ 1.60 for $n$ = 9 -- 11 (equation \ref{eqn:T_less_L}). Higher values of $n$ do not change $s$ substantially. Therefore constant luminosity requires that the density of the CSM falls off slower than a steady wind. If interpreted as a  change in the mass-loss rate, this implies that the mass loss rate of the progenitor star was decreasing as it approached core-collapse, which is not generally expected. Alternatively, it could be interpreted as a change in the wind velocity, implying an increasing wind velocity as the star approached core-collapse.}

{The X-ray lightcurve of SN 1957D shows a decreasing trend, but with large error bars. The power-law fit is consistent with almost no decrease. The radio light curve of SN 1957D may help to better interpret the X-ray evolution. L12 show that the radio flux from SN 1957D was decreasing at close to the rate expected from a steady wind until around 1990, but after 1990 the radio light curve steepened, and began declining at a much faster rate $L_X \propto$ $t^{-4}$. In the \citet{Chevalier1982} model, the radio emission also arises from the interaction of the supernova shock with the progenitor wind. Given the steeply decreasing radio light curve of SN 1957D, it is more likely that the X-ray luminosity is also declining, rather than remaining constant with time.}

It should be pointed out that the prior discussion assumes the validity of the self-similar solution, which is not necessarily the case. In the \citet{Chevalier1982} model, the ejecta density flattens close to the center, and the reverse shock interaction with the flat part of the density profile is no longer described by the self-similar solution. It is possible that after a few decades, the reverse shock in SN 1957D could have begun to interact with the flat part of the profile. On the other hand, after 340 years, the reverse shock in the SNR Cas A is still quite far out in the wind. The numerous caveats and uncertainties in the preceding discussion make clear the need for extended follow-up, at all wavelengths, of old SNe/young SNRs.

Without additional data, it is not easy to estimate whether the emission arises from forward or reverse shocked material, given the low counts and absence of decipherable spectral lines. Since higher temperatures would be expected behind the forward shock, those remnants which show a high temperature, $> 2$ keV, are more likely to have emission arising at the forward shock. For those with low electron temperatures $< 1 $ keV, the emission could be from either shock, since the forward shock too may have slowed down after several decades, or the electron temperature may be much lower than the proton temperature at the collisionless shock. Conversely, the emission could have contributions from both forward and reverse shocks, which are difficult to disentangle.

\citet{sn1941c_optical} have detected optical emission from SN 1941C. The dominance of the [OI] and [OIII] lines, coupled with the absence of detectable H$\alpha$ emission, has been interpreted as an indication that the optical emission originates from the reverse-shocked ejecta, similar to the emission from Cas A. The redshifted emission, extending up to 4400 km s$^{-1}$, dominates the fainter blue-shifted emission, which extends out to velocities of 2200 km s$^{-1}$. Given that the electron temperature may lag the ion temperature, the shock velocity derived from the electron temperature could be consistent with the velocities measured from the forbidden lines of oxygen.  The estimated high X-ray temperature of 3.70 keV would suggest origin in the forward-shocked material, consistent with the blue-shifted velocities of up to -2200 km s$^{-1}$, although it is not possible to rule out an ejecta origin. Conversely, the optical and X-ray emission may originate from different shocks.

\begin{figure}
    \centering
    \includegraphics[width=\linewidth]{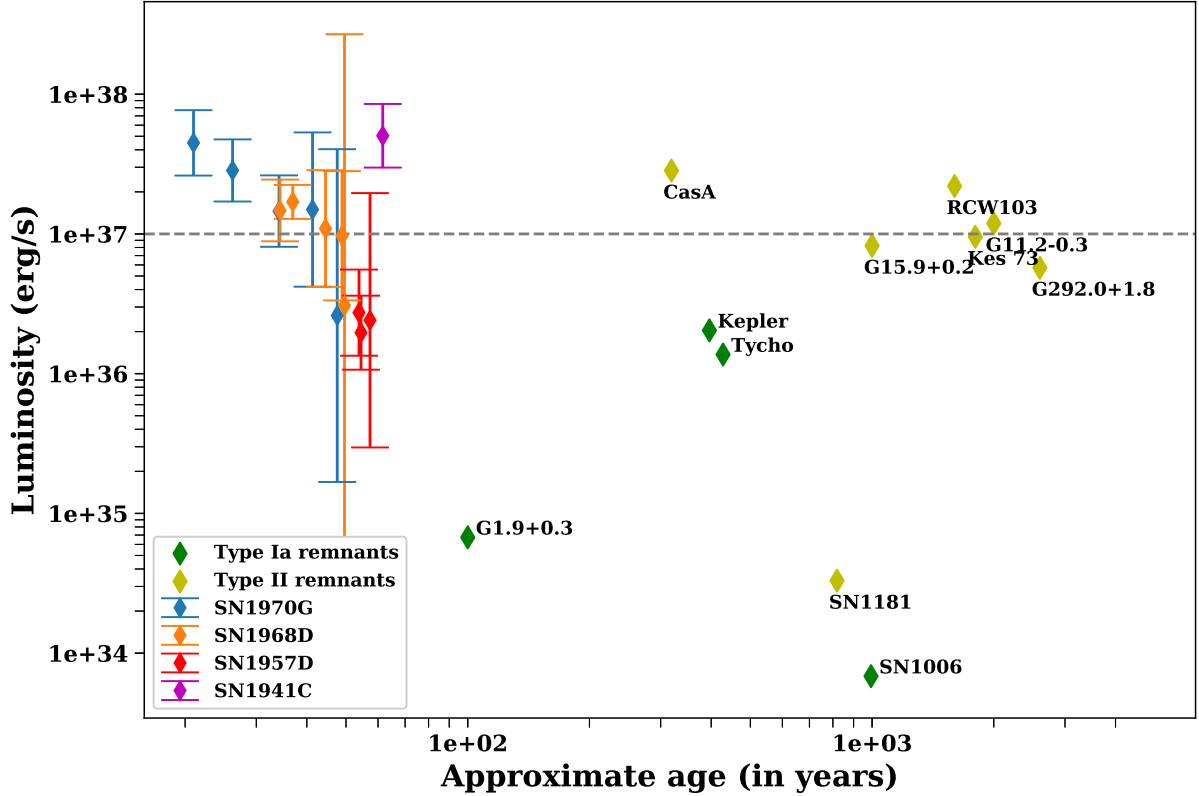}
    \caption{The 0.3 - 8 keV band luminosities of SN 1970G, SN 1968D,
      SN 1957D and SN 1941C, plotted together with the 0.3 - 8 keV band
      luminosities of some Galactic Type Ia and core-collapse
      remnants. Upper limits have been excluded to make the plot more
      readable. }
    \label{fig:all_Lx_snr}
\end{figure}

The 0.3 - 8 keV band X-ray luminosities of SN 1970G, SN 1968D, SN 1957D and SN 1941C are plotted together with the X-ray luminosities of various Galactic remnants in Figure \ref{fig:all_Lx_snr} (see Table \ref{tab:SNR}). {The luminosity of the supernovae considered here appears to decrease with time, albeit with large uncertainties in the powerlaw index. The majority of X-ray supernovae show a similar tendency of decreasing luminosity with time (see for example Figure 1 in \citet{Dwarkadas2014})}. If these old supernovae continue to {evolve} in luminosity at the present rate, or even at a lower rate, then within $\sim$ 300 years they should be significantly fainter than the brightest observed core-collapse SNRs. However, by that time we would also expect the SNe to have swept-up mass several times their ejecta mass, and therefore to have reached the Sedov phase. 

In the Sedov stage, the reverse shock has reached the center and dissipated, and there is only a single forward-propagating shock. The self-similar solution for the radius of the supernova is
\begin{eqnarray}
    r &\propto& t^{2/(5-s)}
    \nonumber \\
    \rho &\propto& r^{-s} \propto t^{-2s/(5-s)}
\end{eqnarray}
By this stage the temperature is likely less than 2.6 $\times$ 10$^7$ K. Following equation \ref{eqn:xray}, the X-ray luminosity evolves as,
\begin{eqnarray}
    L_X &\propto& t^{-4s/(5-s)}\left(\frac{r}{t}\right)^{-0.96}r^3
    \nonumber \\
    L_X &\propto& t^{(8.88-4.96s)/(5-s)}
\end{eqnarray}
We would expect that the remnant is expanding into a constant density interstellar medium, i.e. s = 0. In this regime L$_X$ $\propto$ t$^{1.78}$; {\em the X-ray emission will increase with time}. The reason is obvious - the density is constant while the volume is continually increasing. Thus we expect that once SNe reach the Sedov stage, their X-ray luminosity should generally show an increase, giving rise to the brighter SNRs. More accurate theoretical calculations \citep{Hamilton1983} show a similar rise in luminosity in the Sedov phase. It is possible that some of these young SNe  will therefore likely resemble the more luminous remnants. Others may continue to decline at the present rate or higher, if they run out of material around them, and may evolve to resemble a lower-luminosity remnant such as SN 1181. The only way to understand this transition and bridge the gap between SNe and SNRs is to continue observing older SNe in X-rays until they fade beyond detection.

\begin{table}
    \centering
    \begin{tabular}{c c c c c}
        \hline
        \colhead{Remnant} & \colhead{Type}\footnote{\citet{TypeII,G1.9+0.3}} & \colhead{Luminosity (10$^{36}$ erg s$^{-1}$)}\footnote{From the Chandra Supernova Remnant Catalog (https://hea-www.harvard.edu/ChandraSNR/index.html)} & \colhead{Age (years)} & \colhead{References} \\
        \hline 
        Tycho & Type Ia & 1.37 & 428 & ... \\
        SN1006 & Type Ia & 6.84 $\times$ 10$^{-3}$ & 994 & ... \\
        G1.9+0.3 & Type Ia & 6.74 $\times$ 10$^{-2}$ & 100 & \citep{Reynolds2008} \\
        Kepler & Type Ia & 2.04 & 396 & ... \\
        Cas A & Type II & 28.4 & 319 & \citep{Cas_A}\\
        SN1181 & Type II & 3.30 $\times$ 10$^{-2}$ & 819 & ... \\
        G292.0+1.8 & Type II & 5.73 & 2600 & \citep{G292.0+1.8}\\
        G15.9+0.2 & Type II & 8.24 & 1000 & \citep{G15.9+0.2}\\
        Kes 73 & Type II & 9.53 & 1800 & \citep{Kes73}\\
        RCW103 & Type II & 22.0 & 1600 & \citep{RCW103}\\
        G11.2-0.3 & Type II & 11.9 & 2000 & \citep{G11.2+0.3}\\
        \hline
    \end{tabular}
    \caption{The supernova remnants used in plotting Figure \ref{fig:all_Lx_snr}}
    \label{tab:SNR}
\end{table}

\section{Conclusion} \label{sec:conclusion}

Several analytic and numerical models \citep{sedov59, Chevalier1982,om88, tm99, Tang2016} describe the evolution of young SNe and supernova remnants, although not necessarily the transition from young
SNe to SNRs. Analytic models by their nature include several approximations, and need to be validated against observations. Unfortunately this is a difficult task to accomplish, given the paucity of young remnants. Young SNe are not followed in X-rays for very long, and tend to fade away after a few decades; remnants of a hundred to a few hundred years are scarce, with the low number not enough to make statistical calculations. In order to gain insight into the evolution of supernovae, and to understand the transition from supernovae to supernova remnants, it is desirable to construct the light curves of supernovae over long periods of time, out to late epochs.  In this paper we have examined archival \textit{Chandra} observations of five of the oldest known X-ray supernovae, SN 1970G, SN 1968D, SN 1959D, SN 1957D and SN 1941C. We were able to construct light curves over multiple epochs for three of these, namely SN 1970G, SN 1968D and SN 1957D. 

The late time luminosities of all five supernovae considered here are similar. The luminosity of SN 1980K, although not calculated here, has been shown by SP08 to be 4 $\times 10^{37}$ erg s$^{-1}$ at an age over 20 years, and by \citet{fridrikssonetal08} to be around 3.2 $\times 10^{37}$ erg s$^{-1}$, which would lie within the same luminosity range. 



{We have been able to reconstruct the lightcurves of SN 1970G, SN 1968D and SN 1957D, and extend them to late times. We find that the lightcurves appear to decline with time, although with large error bars. Declining lightcurves are characteristic of most young SNe \citep{Dwarkadas_2012, drrb16}}. The lightcurve of SN 1970G does not exhibit a sudden rise at the 2011 epoch as discussed by \citet{midlife_crisis}, {nor at subsequent epochs}, and thus does not favor the model of an emerging pulsar wind nebula.

The X-ray luminosity of these SNe will likely decrease in the ejecta-dominated phase, as is seen for almost all X-ray SNe.  However, as shown in Section \ref{subsec:light_curves} it should begin to increase once the supernovae reach the Sedov stage. Thus the old SN population we see could potentially form the brighter SNRs that are observed. Although it is also possible that the more luminous remnants could arise from a still brighter population of old SNe, that begs the question of why this bright population of old SNe has not been detected when fainter ones have.

Studies of the transition of a SN to a remnant in the X-rays are very few. There is considerable competition for X-ray time, and TACs are reluctant to allocate time to objects that do not promise `revolutionary' science. Unfortunately, this results in few resolved observations of late-time SNe, which would require large allocations of time. We urge TACs to allocate more observations of late-time SNe in order to better decipher fundamental ideas of SN and shock evolution, and the energy radiated not only at X-ray wavelengths, but at other wavelengths including radio, optical and infra-red.

\acknowledgements We thank the anonymous referee for a very comprehensive reading of the paper, and for their comments and suggestions which significantly helped to improve the paper. This work is supported by a NASA Astrophysics Data Analayis program grant NNX14AR63G, and by NSF grant 1911061, awarded to PI V.~Dwarkadas at the University of Chicago.

\software{CIAO (version 4.10) \citep{ciao}, Sherpa \citep{sherpa}}

\facilities{CXO, CDS}

\bibliographystyle{aasjournal}
\bibliography{refs}

\begin{thebibliography}{}
\expandafter\ifx\csname natexlab\endcsname\relax\def\natexlab#1{#1}\fi
\providecommand{\url}[1]{\href{#1}{#1}}
\providecommand{\dodoi}[1]{doi:~\href{http://doi.org/#1}{\nolinkurl{#1}}}
\providecommand{\doeprint}[1]{\href{http://ascl.net/#1}{\nolinkurl{http://ascl.net/#1}}}
\providecommand{\doarXiv}[1]{\href{https://arxiv.org/abs/#1}{\nolinkurl{https://arxiv.org/abs/#1}}}

\bibitem[{{Bochenek} {et~al.}(2018){Bochenek}, {Dwarkadas}, {Silverman}, {Fox},
  {Chevalier}, {Smith}, \& {Filippenko}}]{bocheneketal18}
{Bochenek}, C.~D., {Dwarkadas}, V.~V., {Silverman}, J.~M., {et~al.} 2018,
  \mnras, 473, 336, \dodoi{10.1093/mnras/stx2029}

\bibitem[{{Borkowski} \& {Reynolds}(2017)}]{Kes73}
{Borkowski}, K.~J., \& {Reynolds}, S.~P. 2017, \apj, 846, 13,
  \dodoi{10.3847/1538-4357/aa830f}

\bibitem[{{Borkowski} {et~al.}(2010){Borkowski}, {Reynolds}, {Green}, {Hwang},
  {Petre}, {Krishnamurthy}, \& {Willett}}]{G1.9+0.3}
{Borkowski}, K.~J., {Reynolds}, S.~P., {Green}, D.~A., {et~al.} 2010, \apj,
  724, L161, \dodoi{10.1088/2041-8205/724/2/L161}

\bibitem[{{Canizares} {et~al.}(1982){Canizares}, {Kriss}, \&
  {Feigelson}}]{SN1980k}
{Canizares}, C.~R., {Kriss}, G.~A., \& {Feigelson}, E.~D. 1982, \apjl, 253,
  L17, \dodoi{10.1086/183728}

\bibitem[{{Cappellari} {et~al.}(2011){Cappellari}, {Emsellem}, {Krajnovi{\'c}},
  {McDermid}, {Scott}, {Verdoes Kleijn}, {Young}, {Alatalo}, {Bacon}, {Blitz},
  {Bois}, {Bournaud}, {Bureau}, {Davies}, {Davis}, {de Zeeuw}, {Duc},
  {Khochfar}, {Kuntschner}, {Lablanche}, {Morganti}, {Naab}, {Oosterloo},
  {Sarzi}, {Serra}, \& {Weijmans}}]{1970G}
{Cappellari}, M., {Emsellem}, E., {Krajnovi{\'c}}, D., {et~al.} 2011, \mnras,
  413, 813, \dodoi{10.1111/j.1365-2966.2010.18174.x}

\bibitem[{{Carter} {et~al.}(1997){Carter}, {Dickel}, \& {Bomans}}]{RCW103}
{Carter}, L.~M., {Dickel}, J.~R., \& {Bomans}, D.~J. 1997, \pasp, 109, 990,
  \dodoi{10.1086/133971}

\bibitem[{{Chandra} {et~al.}(2009){Chandra}, {Dwarkadas}, {Ray}, {Immler}, \&
  {Pooley}}]{sn1993j}
{Chandra}, P., {Dwarkadas}, V.~V., {Ray}, A., {Immler}, S., \& {Pooley}, D.
  2009, \apj, 699, 388, \dodoi{10.1088/0004-637X/699/1/388}

\bibitem[{{Chevalier}(1982)}]{Chevalier1982}
{Chevalier}, R.~A. 1982, \apj, 258, 790, \dodoi{10.1086/160126}

\bibitem[{{Chevalier} \& {Fransson}(2017)}]{Chevalier2017}
{Chevalier}, R.~A., \& {Fransson}, C. 2017, {Thermal and Non-thermal Emission
  from Circumstellar Interaction}, 875, \dodoi{10.1007/978-3-319-21846-5_34}

\bibitem[{{Cowan} \& {Branch}(1983)}]{cowanetal83}
{Cowan}, J.~J., \& {Branch}, D. 1983, in \baas, Vol.~15, 954

\bibitem[{{Detre}(1974)}]{detre74}
{Detre}, L. 1974, in Astrophysics and Space Science Library, Vol.~45,
  Supernovae and Supernova Remnants, ed. C.~B. {Cosmovici}, 51,
  \dodoi{10.1007/978-94-010-2166-1_5}

\bibitem[{{Detre} \& {Lovas}(1970)}]{1970g_discovery}
{Detre}, L., \& {Lovas}, M. 1970, \iaucirc, 2269, 1

\bibitem[{{Dickey} \& {Lockman}(1990)}]{DL_nH}
{Dickey}, J.~M., \& {Lockman}, F.~J. 1990, \araa, 28, 215,
  \dodoi{10.1146/annurev.aa.28.090190.001243}

\bibitem[{{Dittmann} {et~al.}(2014){Dittmann}, {Soderberg}, {Chomiuk},
  {Margutti}, {Goss}, {Milisavljevic}, \& {Chevalier}}]{midlife_crisis}
{Dittmann}, J.~A., {Soderberg}, A.~M., {Chomiuk}, L., {et~al.} 2014, \apj, 788,
  38, \dodoi{10.1088/0004-637X/788/1/38}

\bibitem[{{Dunlap}(1968)}]{1968d_discovery1}
{Dunlap}, J.~R. 1968, \iaucirc, 2057

\bibitem[{{Dwarkadas}(2014)}]{Dwarkadas2014}
{Dwarkadas}, V.~V. 2014, \mnras, 440, 1917, \dodoi{10.1093/mnras/stu347}

\bibitem[{{Dwarkadas} \& {Chevalier}(1998)}]{dc98}
{Dwarkadas}, V.~V., \& {Chevalier}, R.~A. 1998, \apj, 497, 807,
  \dodoi{10.1086/305478}

\bibitem[{{Dwarkadas} \& {Gruszko}(2012)}]{Dwarkadas_2012}
{Dwarkadas}, V.~V., \& {Gruszko}, J. 2012, \mnras, 419, 1515,
  \dodoi{10.1111/j.1365-2966.2011.19808.x}

\bibitem[{{Dwarkadas} {et~al.}(2016){Dwarkadas}, {Romero-Ca{\~n}izales},
  {Reddy}, \& {Bauer}}]{drrb16}
{Dwarkadas}, V.~V., {Romero-Ca{\~n}izales}, C., {Reddy}, R., \& {Bauer}, F.~E.
  2016, \mnras, 462, 1101, \dodoi{10.1093/mnras/stw1717}

\bibitem[{{Eck} {et~al.}(2002){Eck}, {Cowan}, \& {Branch}}]{ecketal02}
{Eck}, C.~R., {Cowan}, J.~J., \& {Branch}, D. 2002, \apj, 573, 306,
  \dodoi{10.1086/340583}

\bibitem[{{Fesen} \& {Weil}(2019)}]{sn1941c_optical}
{Fesen}, R., \& {Weil}, K. 2019, arXiv e-prints, arXiv:1910.07723.
\newblock \doarXiv{1910.07723}

\bibitem[{{Fesen} {et~al.}(2006){Fesen}, {Hammell}, {Morse}, {Chevalier},
  {Borkowski}, {Dopita}, {Gerardy}, {Lawrence}, {Raymond}, \& {van den
  Bergh}}]{Cas_A}
{Fesen}, R.~A., {Hammell}, M.~C., {Morse}, J., {et~al.} 2006, \apj, 645, 283,
  \dodoi{10.1086/504254}

\bibitem[{{Fransson} {et~al.}(1996){Fransson}, {Lundqvist}, \&
  {Chevalier}}]{Fransson1996}
{Fransson}, C., {Lundqvist}, P., \& {Chevalier}, R.~A. 1996, \apj, 461, 993,
  \dodoi{10.1086/177119}

\bibitem[{{Freeman} {et~al.}(2001){Freeman}, {Doe}, \&
  {Siemiginowska}}]{sherpa}
{Freeman}, P., {Doe}, S., \& {Siemiginowska}, A. 2001, in Society of
  Photo-Optical Instrumentation Engineers (SPIE) Conference Series, Vol. 4477,
  \procspie, ed. J.-L. {Starck} \& F.~D. {Murtagh}, 76--87,
  \dodoi{10.1117/12.447161}

\bibitem[{{Fridriksson} {et~al.}(2008){Fridriksson}, {Homan}, {Lewin}, {Kong},
  \& {Pooley}}]{fridrikssonetal08}
{Fridriksson}, J.~K., {Homan}, J., {Lewin}, W. H.~G., {Kong}, A. K.~H., \&
  {Pooley}, D. 2008, \apjs, 177, 465, \dodoi{10.1086/588817}

\bibitem[{{Fruscione} {et~al.}(2006){Fruscione}, {McDowell}, {Allen},
  {Brickhouse}, {Burke}, {Davis}, {Durham}, {Elvis}, {Galle}, {Harris},
  {Huenemoerder}, {Houck}, {Ishibashi}, {Karovska}, {Nicastro}, {Noble},
  {Nowak}, {Primini}, {Siemiginowska}, {Smith}, \& {Wise}}]{ciao}
{Fruscione}, A., {McDowell}, J.~C., {Allen}, G.~E., {et~al.} 2006, in Society
  of Photo-Optical Instrumentation Engineers (SPIE) Conference Series, Vol.
  6270, Society of Photo-Optical Instrumentation Engineers (SPIE) Conference
  Series, 62701V, \dodoi{10.1117/12.671760}

\bibitem[{{Gil de Paz} {et~al.}(2007){Gil de Paz}, {Boissier}, {Madore},
  {Seibert}, {Joe}, {Boselli}, {Wyder}, {Thilker}, {Bianchi}, {Rey}, {Rich},
  {Barlow}, {Conrow}, {Forster}, {Friedman}, {Martin}, {Morrissey}, {Neff},
  {Schiminovich}, {Small}, {Donas}, {Heckman}, {Lee}, {Milliard}, {Szalay}, \&
  {Yi}}]{1959D}
{Gil de Paz}, A., {Boissier}, S., {Madore}, B.~F., {et~al.} 2007, \apj, 173,
  185, \dodoi{10.1086/516636}

\bibitem[{{Gonzalez} \& {Safi-Harb}(2003)}]{G292.0+1.8}
{Gonzalez}, M., \& {Safi-Harb}, S. 2003, \apj, 583, L91, \dodoi{10.1086/368122}

\bibitem[{{Gottesman} {et~al.}(1972){Gottesman}, {Broderick}, {Brown},
  {Balick}, \& {Palmer}}]{gottesmanetal72}
{Gottesman}, S.~T., {Broderick}, J.~J., {Brown}, R.~L., {Balick}, B., \&
  {Palmer}, P. 1972, \apj, 174, 383, \dodoi{10.1086/151497}

\bibitem[{{Hamilton} {et~al.}(1983){Hamilton}, {Sarazin}, \&
  {Chevalier}}]{Hamilton1983}
{Hamilton}, A.~J.~S., {Sarazin}, C.~L., \& {Chevalier}, R.~A. 1983, \apjs, 51,
  115, \dodoi{10.1086/190841}

\bibitem[{{Humason}(1959)}]{1959d_discovery}
{Humason}, M.~L. 1959, \iaucirc, 1682, 1

\bibitem[{{Immler} \& {Kuntz}(2005)}]{Immler_2005}
{Immler}, S., \& {Kuntz}, K.~D. 2005, \apj, 632, L99, \dodoi{10.1086/497910}

\bibitem[{{Immler} \& {Lewin}(2003)}]{Immler2003}
{Immler}, S., \& {Lewin}, W.~H.~G. 2003, {X-Ray Supernovae}, ed. K.~{Weiler},
  Vol. 598, 91--111, \dodoi{10.1007/3-540-45863-8_7}

\bibitem[{{Jones}(1941)}]{1941c_discovery}
{Jones}, R. 1941, \iaucirc, 866, 1

\bibitem[{{Kent}(1985)}]{Kent85}
{Kent}, S.~M. 1985, \apjs, 59, 115, \dodoi{10.1086/191066}

\bibitem[{{Kowal} \& {Sargent}(1971)}]{1957d_discovery}
{Kowal}, C.~T., \& {Sargent}, W.~L.~W. 1971, \aj, 76, 756,
  \dodoi{10.1086/111193}

\bibitem[{{Kraft} {et~al.}(1991){Kraft}, {Burrows}, \& {Nousek}}]{kraft1991}
{Kraft}, R.~P., {Burrows}, D.~N., \& {Nousek}, J.~A. 1991, \apj, 374, 344,
  \dodoi{10.1086/170124}

\bibitem[{{Lelli} {et~al.}(2016){Lelli}, {McGaugh}, \& {Schombert}}]{1968D}
{Lelli}, F., {McGaugh}, S.~S., \& {Schombert}, J.~M. 2016, \apj, 816, L14,
  \dodoi{10.3847/2041-8205/816/1/L14}

\bibitem[{{Long} {et~al.}(1989){Long}, {Blair}, \& {Krzeminski}}]{longetal89}
{Long}, K.~S., {Blair}, W.~P., \& {Krzeminski}, W. 1989, \apjl, 340, L25,
  \dodoi{10.1086/185430}

\bibitem[{{Long} {et~al.}(2014){Long}, {Kuntz}, {Blair}, {Godfrey},
  {Plucinsky}, {Soria}, {Stockdale}, \& {Winkler}}]{Long_2014}
{Long}, K.~S., {Kuntz}, K.~D., {Blair}, W.~P., {et~al.} 2014, \apjs, 212, 21,
  \dodoi{10.1088/0067-0049/212/2/21}

\bibitem[{{Long} {et~al.}(2012){Long}, {Blair}, {Godfrey}, {Kuntz},
  {Plucinsky}, {Soria}, {Stockdale}, {Whitmore}, \& {Winkler}}]{Long_2012}
{Long}, K.~S., {Blair}, W.~P., {Godfrey}, L.~E.~H., {et~al.} 2012, \apj, 756,
  18, \dodoi{10.1088/0004-637X/756/1/18}

\bibitem[{{Lopez} {et~al.}(2009){Lopez}, {Ramirez-Ruiz}, {Badenes},
  {Huppenkothen}, {Jeltema}, \& {Pooley}}]{TypeII}
{Lopez}, L.~A., {Ramirez-Ruiz}, E., {Badenes}, C., {et~al.} 2009, \apj, 706,
  L106, \dodoi{10.1088/0004-637X/706/1/L106}

\bibitem[{{Marinova} \& {Jogee}(2007)}]{1941C}
{Marinova}, I., \& {Jogee}, S. 2007, \apj, 659, 1176, \dodoi{10.1086/512355}

\bibitem[{{Matzner} \& {McKee}(1999)}]{mm99}
{Matzner}, C.~D., \& {McKee}, C.~F. 1999, \apj, 510, 379,
  \dodoi{10.1086/306571}

\bibitem[{{Ostriker} \& {McKee}(1988)}]{om88}
{Ostriker}, J.~P., \& {McKee}, C.~F. 1988, Reviews of Modern Physics, 60, 1,
  \dodoi{10.1103/RevModPhys.60.1}

\bibitem[{{Reynolds} {et~al.}(2008){Reynolds}, {Borkowski}, {Green}, {Hwang},
  {Harrus}, \& {Petre}}]{Reynolds2008}
{Reynolds}, S.~P., {Borkowski}, K.~J., {Green}, D.~A., {et~al.} 2008, \apjl,
  680, L41, \dodoi{10.1086/589570}

\bibitem[{{Reynolds} {et~al.}(2006){Reynolds}, {Borkowski}, {Hwang}, {Harrus},
  {Petre}, \& {Dubner}}]{G15.9+0.2}
{Reynolds}, S.~P., {Borkowski}, K.~J., {Hwang}, U., {et~al.} 2006, \apj, 652,
  L45, \dodoi{10.1086/510066}

\bibitem[{{Ross} \& {Dwarkadas}(2017)}]{rd17}
{Ross}, M., \& {Dwarkadas}, V.~V. 2017, \aj, 153, 246,
  \dodoi{10.3847/1538-3881/aa6d50}

\bibitem[{{Schlegel}(1994)}]{schlegel94}
{Schlegel}, E.~M. 1994, \aj, 108, 1893, \dodoi{10.1086/117202}

\bibitem[{{Schlegel}(1995)}]{Schlegel1995}
---. 1995, Reports on Progress in Physics, 58, 1375,
  \dodoi{10.1088/0034-4885/58/11/001}

\bibitem[{{Sedov}(1959)}]{sedov59}
{Sedov}, L.~I. 1959, {Similarity and Dimensional Methods in Mechanics}

\bibitem[{{Soria} \& {Perna}(2008)}]{Soria_2008}
{Soria}, R., \& {Perna}, R. 2008, \apj, 683, 767, \dodoi{10.1086/589995}

\bibitem[{{Stockdale} {et~al.}(1998){Stockdale}, {Romanishin}, \&
  {Cowan}}]{Stockdale1998}
{Stockdale}, C.~J., {Romanishin}, W., \& {Cowan}, J.~J. 1998, \apjl, 508, L33,
  \dodoi{10.1086/311706}

\bibitem[{{Tang} \& {Chevalier}(2017)}]{Tang2016}
{Tang}, X., \& {Chevalier}, R.~A. 2017, \mnras, 465, 3793,
  \dodoi{10.1093/mnras/stw2978}

\bibitem[{{Truelove} \& {McKee}(1999)}]{tm99}
{Truelove}, J.~K., \& {McKee}, C.~F. 1999, \apjs, 120, 299,
  \dodoi{10.1086/313176}

\bibitem[{{Tully} {et~al.}(2016){Tully}, {Courtois}, \& {Sorce}}]{1957D}
{Tully}, R.~B., {Courtois}, H.~M., \& {Sorce}, J.~G. 2016, \aj, 152, 50,
  \dodoi{10.3847/0004-6256/152/2/50}

\bibitem[{{Vasisht} {et~al.}(1996){Vasisht}, {Aoki}, {Dotani}, {Kulkarni}, \&
  {Nagase}}]{G11.2+0.3}
{Vasisht}, G., {Aoki}, T., {Dotani}, T., {Kulkarni}, S.~R., \& {Nagase}, F.
  1996, \apj, 456, L59, \dodoi{10.1086/309854}

\bibitem[{{Wenger} {et~al.}(2000){Wenger}, {Ochsenbein}, {Egret}, {Dubois},
  {Bonnarel}, {Borde}, {Genova}, {Jasniewicz}, {Lalo{\"e}}, {Lesteven}, \&
  {Monier}}]{SIMBAD}
{Wenger}, M., {Ochsenbein}, F., {Egret}, D., {et~al.} 2000, \aaps, 143, 9,
  \dodoi{10.1051/aas:2000332}

\bibitem[{{Wild}(1968)}]{1968d_discovery2}
{Wild}, P. 1968, \iaucirc, 2057, 1

\end{thebibliography}

\end{document}